\documentclass[fleqn,usenatbib]{mnras}

\usepackage{newtxtext,newtxmath}

\usepackage[T1]{fontenc}
\usepackage{ae,aecompl}

\usepackage{graphicx}	
\usepackage{amsmath}	
\usepackage{enumerate}
\usepackage{color}
\usepackage{graphicx}
\usepackage{multicol}
\usepackage{savesym}
\usepackage{color}
\usepackage{tipa}
\usepackage{csvsimple}
\usepackage{arydshln}
\usepackage{arydshln}

\title[ULX bubble in NGC\,5585]
  {The ultraluminous X-ray source bubble in NGC\,5585}

\author[R. Soria et al.]
{R.~Soria$^{1,2}$ \thanks{Email: rsoria@nao.cas.cn (RS)}, 
M.W.~Pakull$^{3}$,
C.~Motch$^{3}$,
J.C.A.~Miller-Jones$^{4}$,
A.D.~Schwope$^{5}$,\newauthor
R.T.~Urquhart$^{6}$,
and M.S.~Ryan$^{4}$\\
\\
$^{1}$College of Astronomy and Space Sciences, University of the Chinese Academy of Sciences, Beijing 100049, China\\
$^{2}$Sydney Institute for Astronomy, School of Physics A28, The University of Sydney, Sydney, NSW 2006, Australia\\
$^{3}$ Universit\'e de Strasbourg, CNRS, Observatoire astronomique, CNRS, UMR 7550,F-67000, Strasbourg, France\\
$^{4}$International Centre for Radio Astronomy Research, Curtin University, GPO Box U1987, Perth, WA 6845, Australia\\
$^{5}$Leibniz-Institut f\"ur Astrophysik Potsdam (AIP), An der Sternwarte 16, 14482, Potsdam, Germany\\
$^{6}$Center for Data Intensive and Time Domain Astronomy, Department of Physics and Astronomy, Michigan State University, East Lansing, MI, USA
\\
}

\date{Accepted 2020 December 3. Received 2020 November 11; in original form 2020 August 17}

\pubyear{2021}

\begin{document}
\label{firstpage}
\pagerange{\pageref{firstpage}--\pageref{lastpage}}
\maketitle

\begin{abstract}
Some ultraluminous X-ray sources (ULXs) are surrounded by collisionally ionized bubbles, larger and more energetic than supernova remnants: they are evidence of the powerful outflows associated with super-Eddington X-ray sources. We illustrate the most recent addition to this class: a huge (350 pc $\times$ 220 pc in diameter) bubble around a ULX in NGC\,5585. We modelled the X-ray properties of the ULX (a broadened-disc source with $L_{\rm X} \approx 2$--4 $\times 10^{39}$ erg s$^{-1}$) from {\it Chandra} and {\it XMM-Newton}, and identified its likely optical counterpart in {\it Hubble Space Telescope} images. We used the Large Binocular Telescope to study the optical emission from the ionized bubble. We show that the line emission spectrum is indicative of collisional ionization. We refine the method for inferring the shock velocity from the width of the optical lines. We derive an average shock velocity $\approx$125 km s$^{-1}$, which corresponds to a dynamical age of $\sim$600,000 years for the bubble, and an average mechanical power $P_{\rm w} \sim 10^{40}$ erg s$^{-1}$; thus, the mechanical power is a few times higher than the current photon luminosity.
With Very Large Array observations, we discovered and resolved a powerful radio bubble with the same size as the optical bubble, and a 1.4-GHz luminosity $\sim$10$^{35}$ erg s$^{-1}$, at the upper end of the luminosity range for this type of source. We explain why ULX bubbles tend to become more radio luminous as they expand while radio supernova remnants tend to fade. 
\end{abstract}


\section{Introduction}
In the process of accretion onto a stellar-mass compact object, the relative output of radiative and mechanical power is a function of accretion rate and of other parameters such as the geometry of the inflow and the nature of the compact object. Typical regimes well studied in Galactic X-ray binaries \citep{fender04,remillard06,fender09} are: (i) the low/hard state, where the kinetic power of a collimated jet increasingly dominates over photon emission, for low accretion rates \citep{fender03}; and (ii) the disk-dominated thermal state (high/soft state), where the jet is quenched \citep{meier01,russell11} and the accretion flow forms a radiatively efficient standard disk. Over the last decade, a lot of observational and theoretical effort has focused on the properties of accretion at even higher accretion rates, above the classical Eddington limit. X-ray binaries in this super-critical regime are generally known as ultraluminous X-ray sources (ULXs:  \citealt{kaaret17,feng11}). One of the hallmark predictions for the super-critical accretion regime is the presence of strong radiatively-driven outflows, launched from the disk surface \citep{king03,poutanen07,dotan11}. Magneto-hydrodynamical simulations \citep{ohsuga11,kawashima12,jiang14,ogawa17,narayan17} suggest that the massive disk wind creates a lower-density funnel along the polar directions; a collimated jet may be launched inside that funnel. The presence of strong outflows has been directly confirmed by X-ray spectroscopic studies of a few nearby ULXs  \citep{pinto16,walton16,pinto17,kosec18}

The accretion models and observations cited above suggest that the kinetic power of outflows from super-critical accretors is of the same order of magnitude as the radiative power. One of the most effective ways to identify the presence of strong outflows and measure or constrain their power is to search for large ($\sim$100 pc) bubbles of shock-ionized gas around the compact object (``ULX bubbles'': \citealt{pakull02,wang02,pakull06,ramsey06,soria10}). The size and expansion velocity of a shock-ionized bubble constrain its kinetic energy \citep{weaver77} and its age, which is a proxy for the duration of the super-critical accretion phase. The flux in diagnostic optical lines (particularly, H$\beta$) is another, independent proxy for the input power that inflates the bubble \citep{dopita95,pakull10}. We have been conducting a long-term program of search, identification and modelling of ULX bubbles in nearby galaxies. In addition to a better understanding of accretion processes in ULXs, modelling the properties of such bubbles provides a template to understand other astrophysical phenomena, such as phases of enhanced nuclear activity in normal galaxies \citep[{\it e.g.},][]{guo12,rampadarath18}, feedback processes in quasars 
\citep[{\it e.g.},][]{king15,tombesi15,parker17}, or the role of X-ray binaries and microquasars in cosmic re-ionization 
\citep[{\it e.g.},][]{mirabel11,fragos13,madau17,douna18}.


In this paper, we present the first detailed study of a huge ($\approx$350 by 220 pc) shock-ionized bubble in the outskirts of the Sd galaxy NGC\,5585 (Figure 1), a member of the M\,101 group \citep{tikhonov15,karachentsev19}, with a star formation rate of $\approx$0.4 $M_{\odot}$ yr$^{-1}$ \citep{james04}.  An accurate distance to NGC\,5585 remains elusive, in the absence of Cepheid or tip-of-the-red-giant-branch measurements. A series of measurements based on the near-infrared Tully-Fisher relation, listed in the NASA/IPAC Extragalactic Database\footnote{https://ned.ipac.caltech.edu}, suggests an average distance of $\approx$ 8 Mpc for a Hubble constant of $\approx$74 km s$^{-1}$ Mpc$^{-1}$ \citep{riess19}. We adopt $d = 8.0$ Mpc in this paper. 

The optical bubble was first noted and studied by \cite{matonick97}; they described it as an ``enormous" and ``particularly interesting" supernova remnant (SNR). Their optical spectrum (taken with the 2.4-m Hiltner telescope at the Michigan-Dartmouth-MIT Observatory in 1994 May) shows that the gas is shock-ionized and suggests velocity broadening for the strongest emission lines. This peculiar ``SNR'' is similar in size and luminosity to another exceptional ``SNR'' also shown in \cite{matonick97}, namely NGC\,7793-S26. It was proposed by \cite{pakull08} that both those ionized bubbles are not SNRs but are instead powered by jets or outflows from a compact object in a super-critical accretion regime. For the NGC\,5585 source, this physical interpretation was supported by the discovery of a ULX (henceforth NGC\,5585 X-1) in the centre of the bubble \citep{pakull08}\footnote{The name ``NGC\,5885'' instead of NGC\,5585 in \cite{pakull08} is a typo.}.
Several other shock-ionized bubbles with similar size ($\sim$100--300 pc) have been identified in nearby galaxies as accretion-powered bubbles. In addition to the already mentioned NGC\,7793-S26 \citep{pakull10,soria10}, the best known are the bubbles around the ULXs NGC\,1313 X-2 \citep{pakull02,weng14}, Holmberg IX X-1 \citep{pakull02,pakull06,moon11}, IC\,342 X-1 \citep{cseh12}.

Here, we investigate the large bubble around NGC\,5585 X-1 in more detail. We collected and studied new and archival data in several bands: in the X-ray band with {\it Chandra} and {\it XMM-Newton}; in the optical band with the {\it Hubble Space Telescope} ({\it HST}) for imaging and the Large Binocular Telescope (LBT\footnote{The LBT is an international collaboration among institutions in the United States, Italy and Germany. LBT Corporation partners are: The University of Arizona on behalf of the Arizona university system; Istituto Nazionale di Astrofisica, Italy; LBT Beteiligungsgesellschaft, Germany, representing the Max-Planck Society, the Astrophysical Institute Potsdam, and Heidelberg University; The Ohio State University, and The Research Corporation, on behalf of The University of Notre Dame, University of Minnesota and University of Virginia.}) for spectroscopy; in the radio band with the Karl G. Jansky Very Large Array (VLA). We will determine the properties of the central point-like source X-1, the shock velocity of the bubble, its dynamical age and mechanical power. We will show that the ionized nebula is indeed one of the most powerful ULX bubbles in the local universe.




\begin{figure}
\vspace{-3.2cm}
\hspace{-0.6cm}
\includegraphics[height=0.76\textwidth]{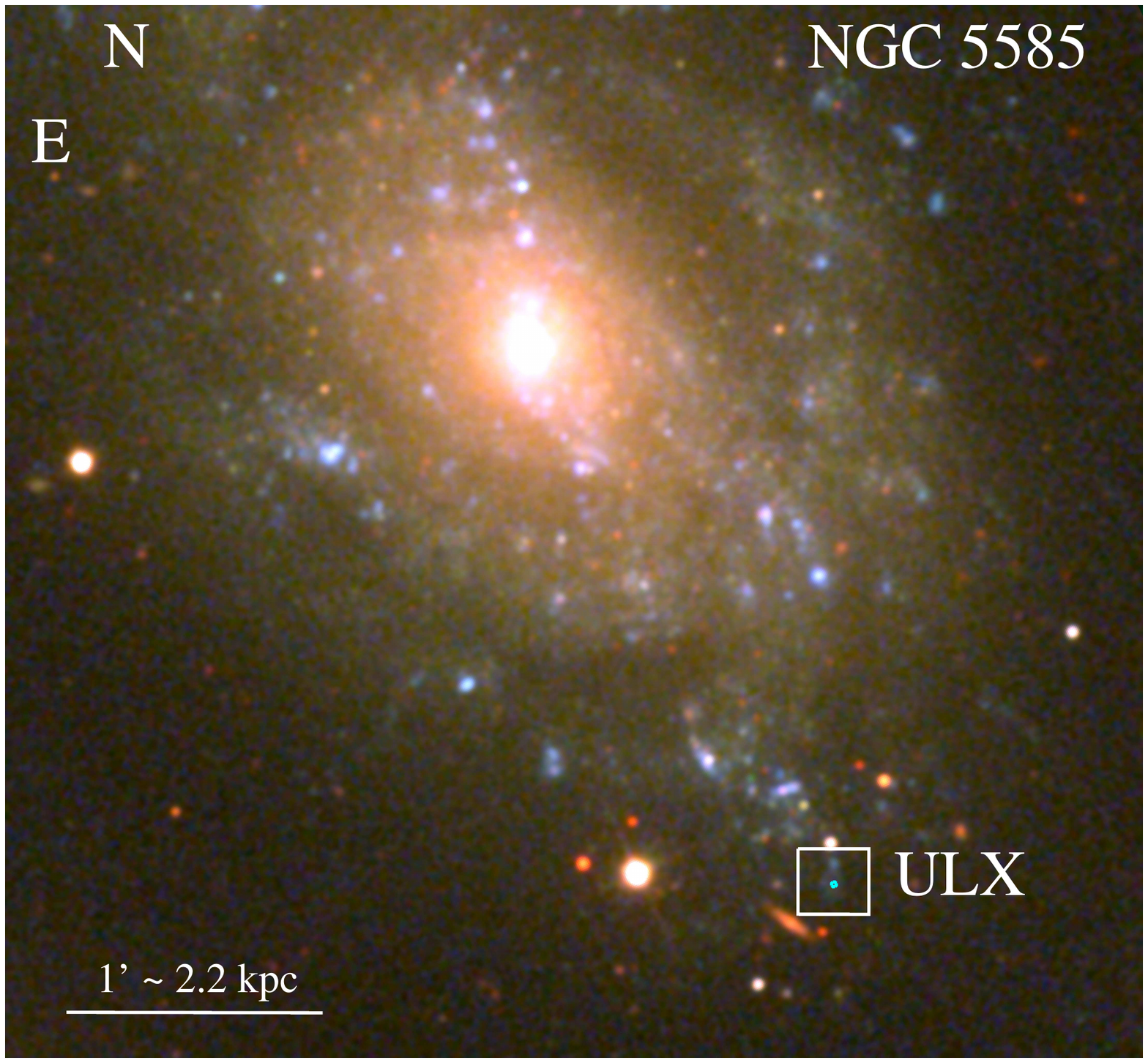}\\[-75pt]
 \caption{Sloan Digital Sky Survey image of NGC\,5585 (red = $i$ band, green = $g$ band, blue = $u$ band), marking the location of the ULX and its bubble. The region inside the white box is displayed in detail in Figure 2.}
  \label{chart}
\end{figure}

\begin{figure*}
\centering
\includegraphics[height=0.305\textwidth]{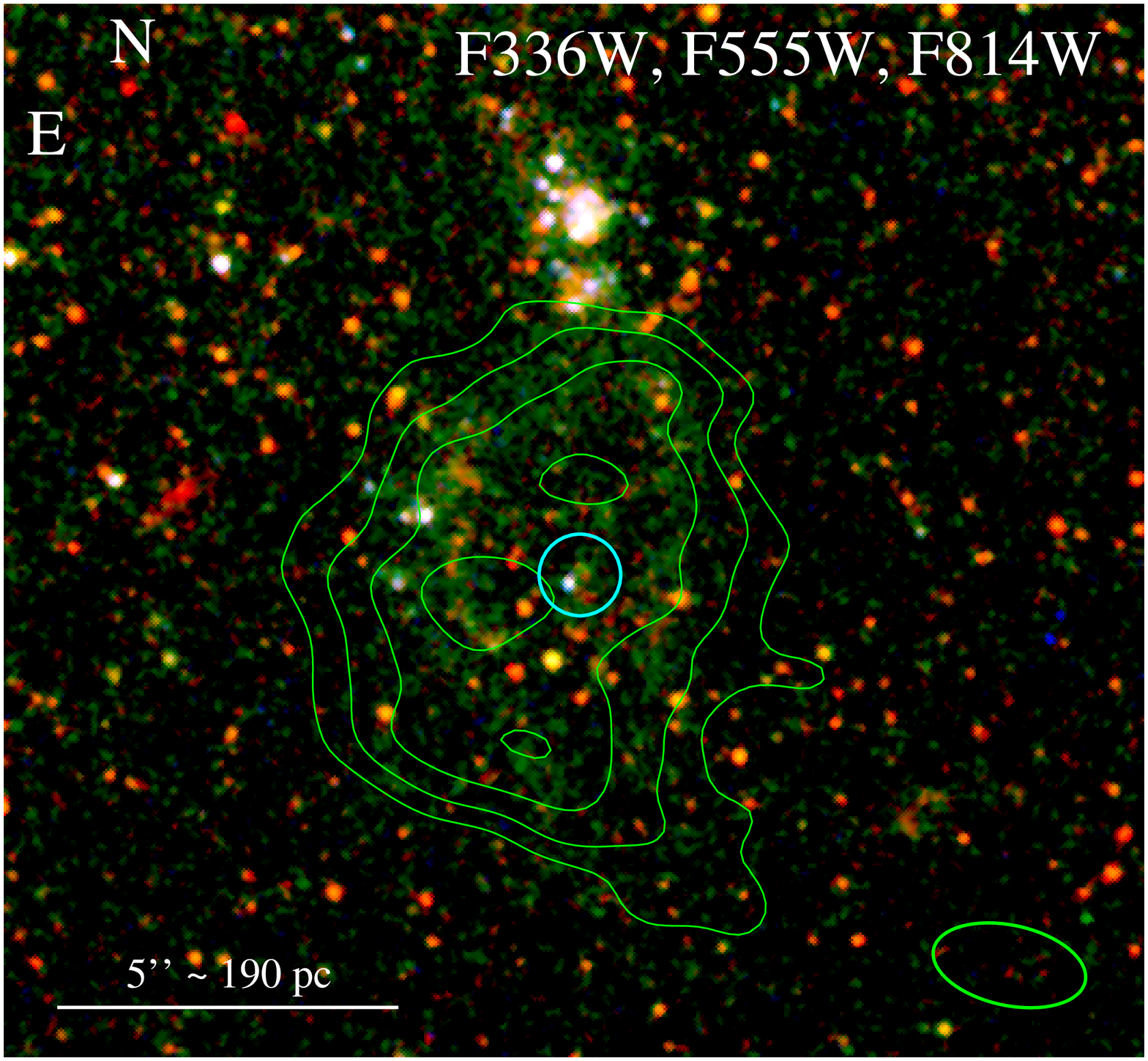}
\includegraphics[height=0.305\textwidth]{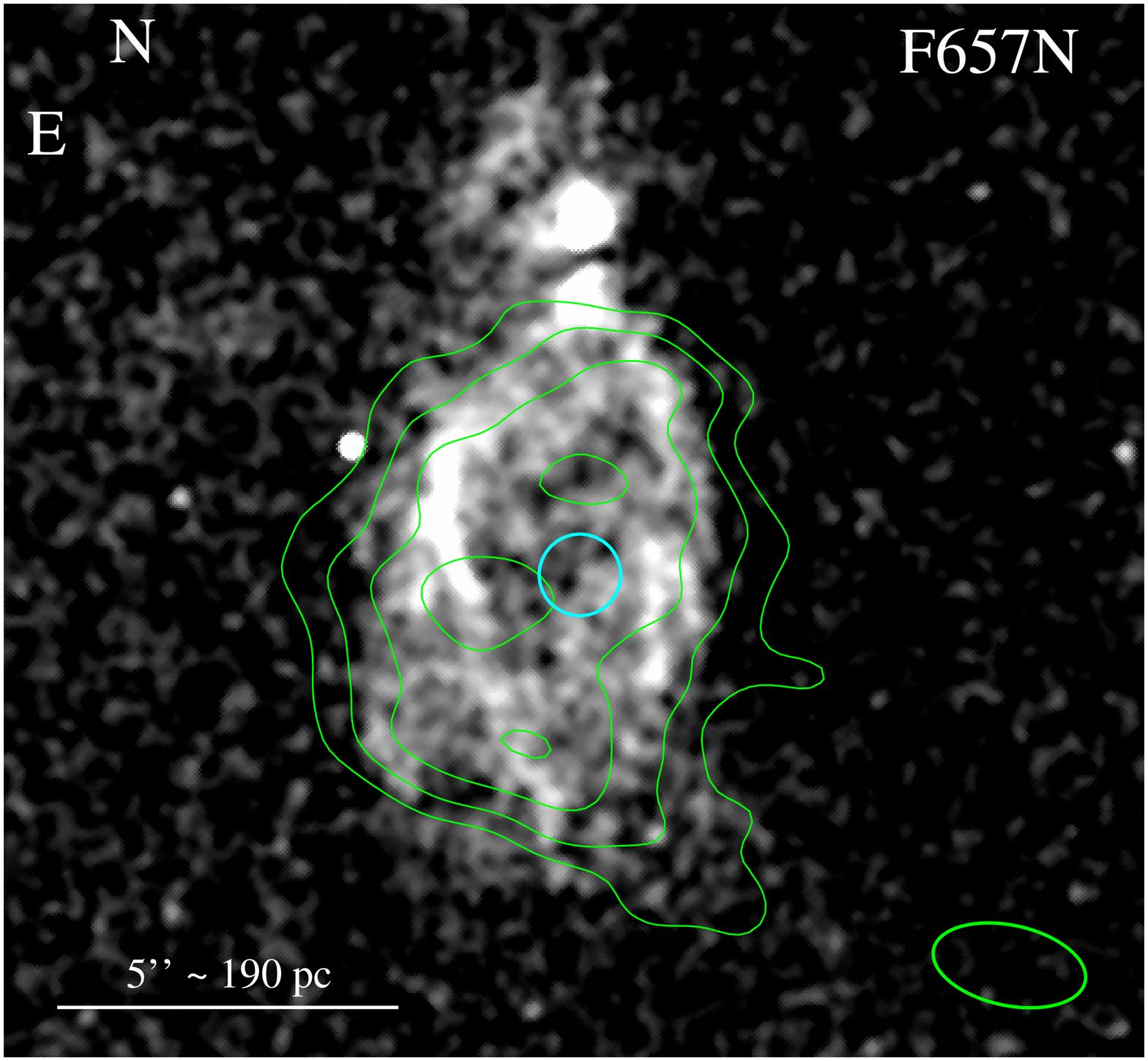}
\includegraphics[height=0.305\textwidth]{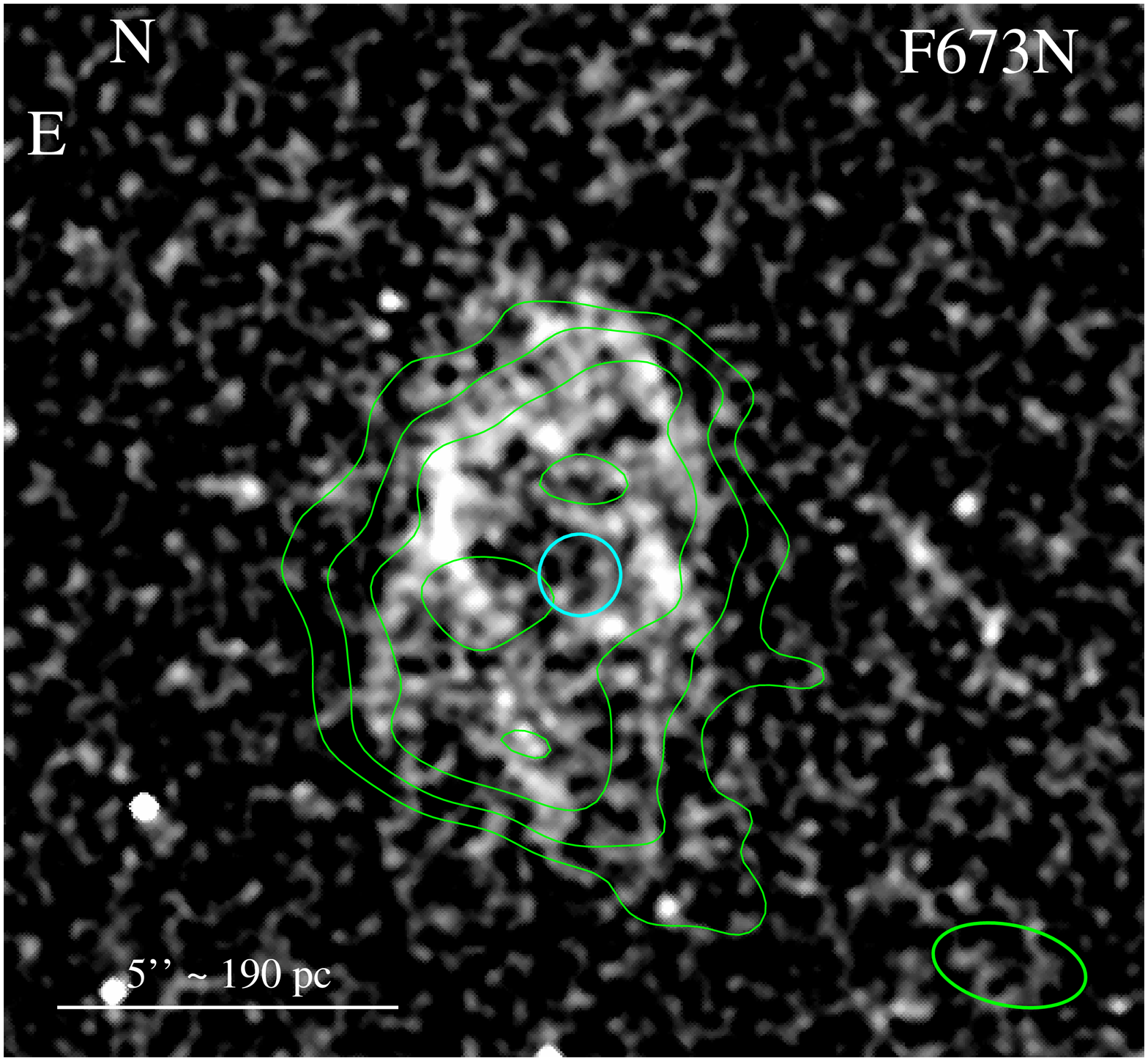}\\
\includegraphics[height=0.305\textwidth]{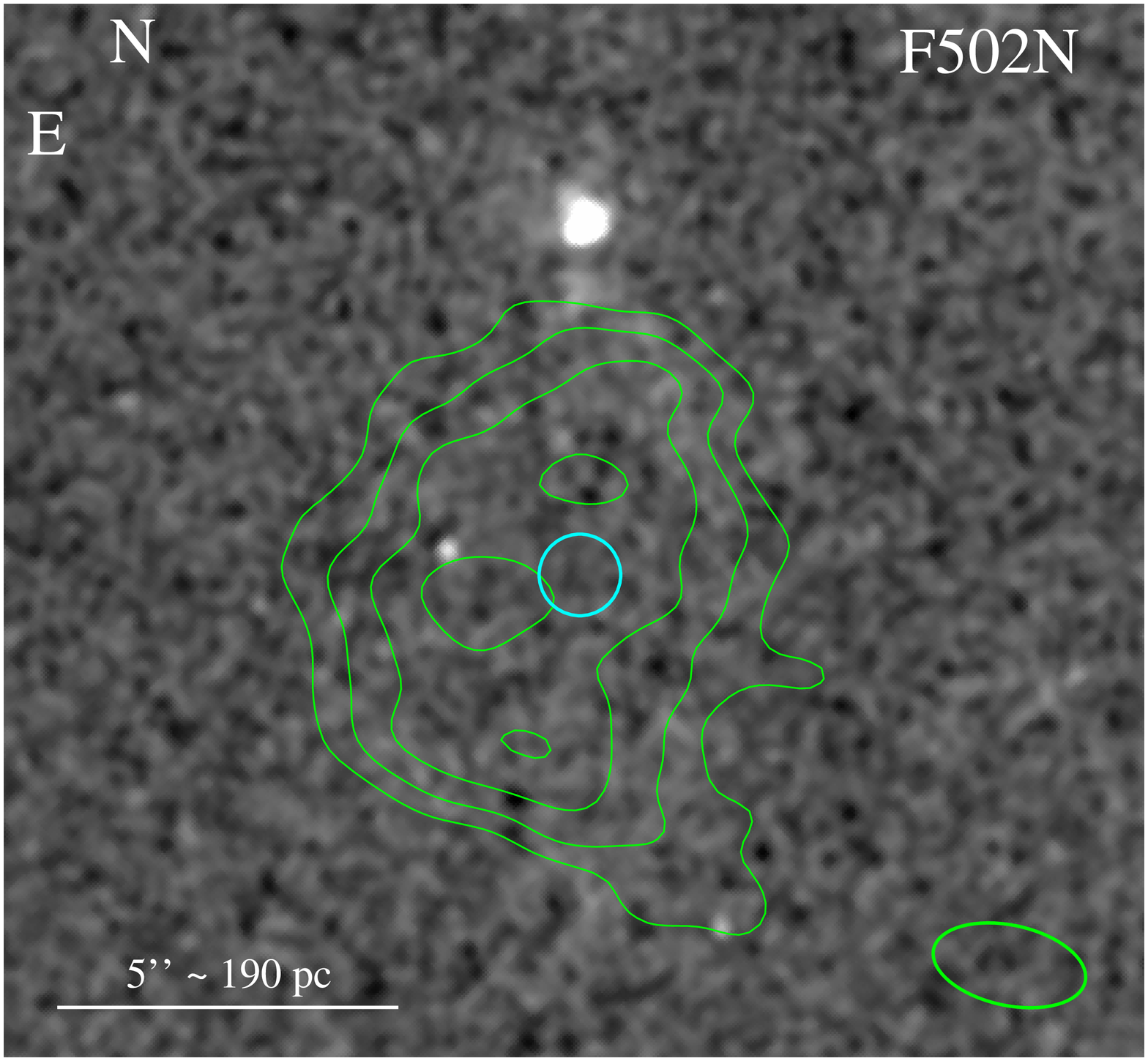}
\includegraphics[height=0.305\textwidth]{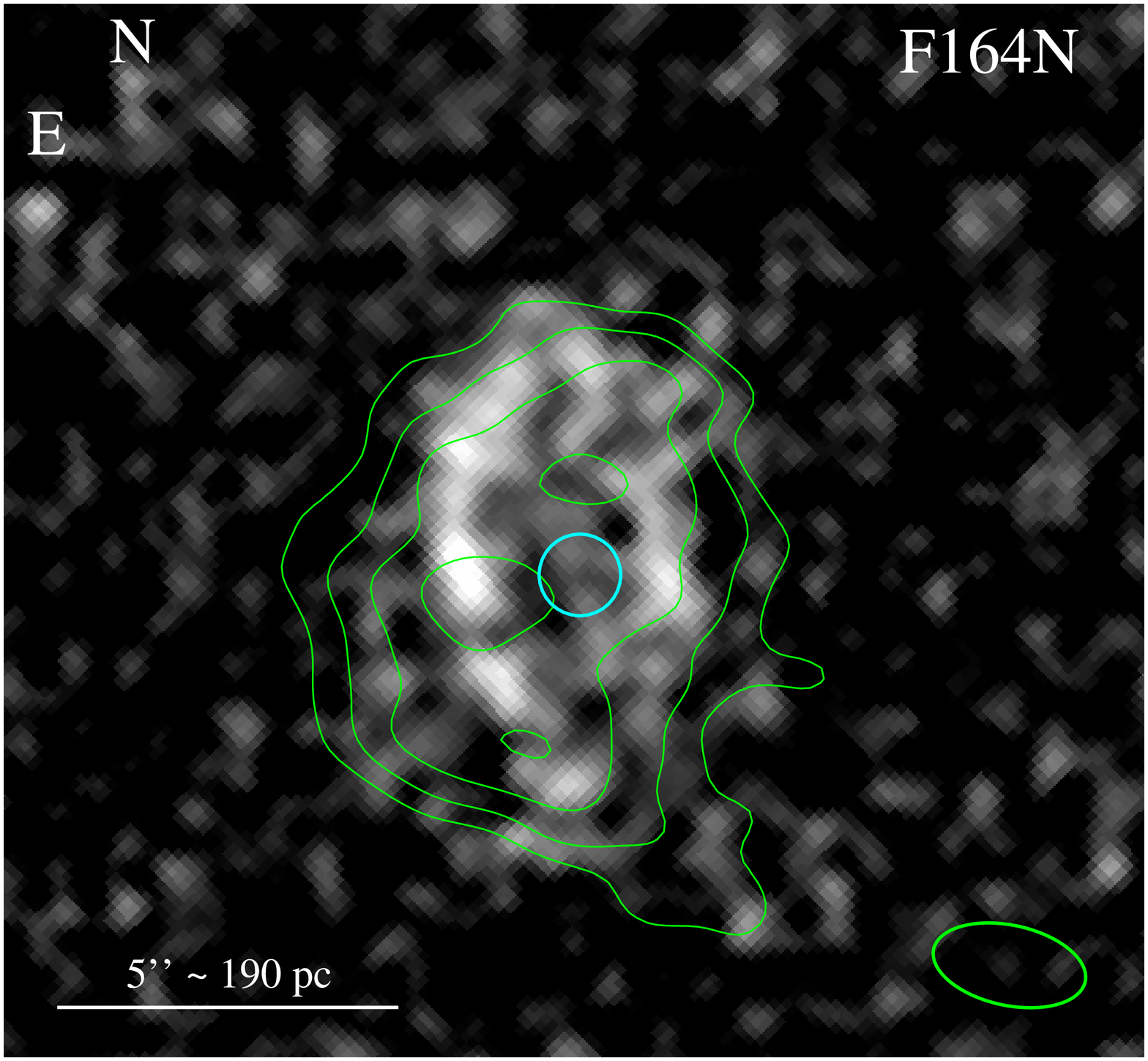}
\includegraphics[height=0.305\textwidth]{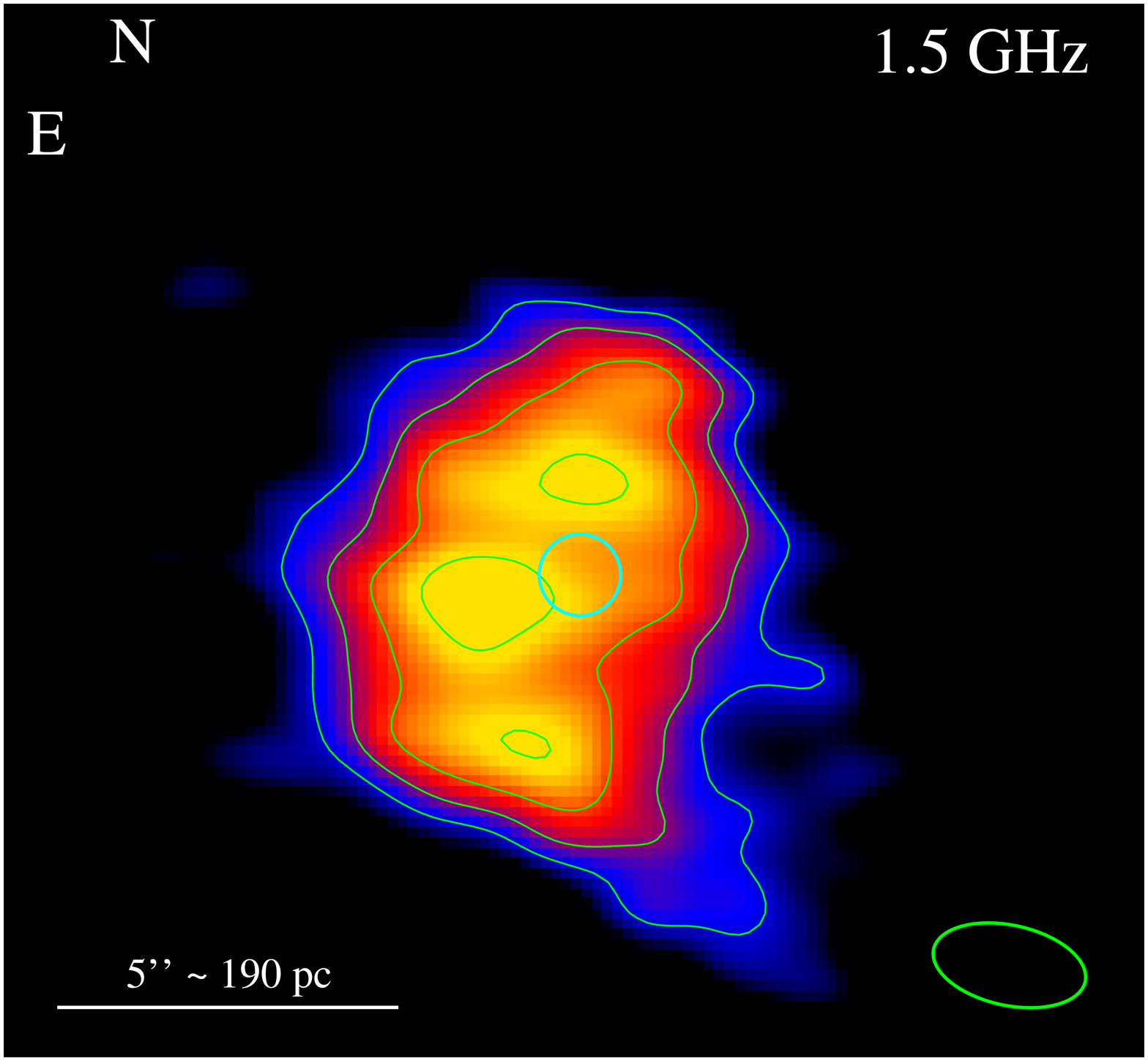}
 \caption{Top row, left panel: {\it HST}/WFC3 image of the field around NGC\,5585 X-1, in the broad-band filters F814W (red), F555W (green) and F336W (blue). The cyan circle marks the location of the ULX (determined from the {\it Chandra} detection), and has a radius of 0$''$.6 (astrometric uncertainty). The green contours show the 1.5-GHz radio emission from the VLA observations; more specifically, they represent flux densities of $2^{n/2}$ times the local rms noise level, with $n =$ 4, 5, 6 and 7 ({\it i.e.}, the lowest contour is a 4-$\sigma$ detection). The green ellipse represents the VLA beam, with major axes of $2''.28 \times 1''.18$, and Position Angle of $78^{\circ}.0$.
 Top row, middle panel: continuum-subtracted image in the {\it HST}/WFC3 F657N filter (the continuum was a linear combination of the F555W and F814W images), which includes H$\alpha$ and [N II]$\lambda \lambda 6548,6584$. The cyan circle, green ellipse and green contours are as in the previous panel.  Top row, right panel: continuum-subtracted image in the {\it HST}/WFC3 F673N filter (the continuum was the same linear combination of F555W and F814W), which covers [S II]$\lambda \lambda 6716,6731$. The cyan circle, green ellipse and green contours are as in the previous panels. Bottom row, left panel: continuum-subtracted image in the {\it HST}/WFC3 F502N filter (F555W was used for the continuum), which covers [O III]$\lambda 5007$. The cyan circle, green ellipse and green contours are as in the previous panels. Bottom row, middle panel: continuum-subtracted image in the {\it HST}/WFC3 F164N filter (F160W was used for the continuum), which covers [Fe II]$\lambda 16440$. The cyan circle, green ellipse and green contours are as in the previous panels. Bottom row, right panel: false-colour 1.5-GHz VLA image with associated flux-density contours and ULX position marked by the cyan circle.}
  \label{panels}
\end{figure*}

\section{Observations and data analysis}

\subsection{{\it Chandra}}

NGC\,5585 was observed by {\it {Chandra}} twice (ObsID 7050 = 2006 August 28, and ObsID 19348 = 2017 May 31), on both occasions only with a 5-ks exposure time (Table 1).  In both observations, the centre of the galaxy was placed at the aimpoint on the S3 chip of the Advanced CCD Imaging Spectrometer array (ACIS). 
We downloaded the public-archive data, and analyzed them with the Chandra Interactive Analysis of Observations ({\sc {ciao}}) software version 6.12 \citep{fruscione06}, with calibration database version 4.9.1. We used the task {\it {chandra\_repro}} to re-build level-2 event files, and we filtered it with {\it dmcopy}. The point-like source NGC\,5585 X-1 is detected in both observations (Section 3.1); we used {\it specextract} to extract  background-subtracted spectra from the two epochs. Finally, we fitted the spectra with {\sc xspec} \citep{arnaud96} version 12.11.0, using the Cash statistics, given the low number of counts.


\subsection{{\it XMM-Newton}}

We observed NGC\,5585 twice with the European Photon Imaging Camera (EPIC), on 2015 June 19 and 2015 June 21, on both occasions with a duration of 39 ks including overheads (live time: 37 ks for MOS1 and MOS2, 32 ks for pn).
We reduced the data with the Science Analysis Software ({\sc sas}) version 17.0.0; we used the {\sc sas} tasks {\it epproc} and {\it emproc} to re-build event files for pn and MOS, respectively. 

We filtered the event files to remove intervals of high particle background. In the first observation, the background level was very low, and the whole exposure can be used for subsequent analysis. In the second observation, the background was substantially higher, and flaring at the beginning and at the end of the observation. We selected good-time-intervals at PI $> 10000$ with a RATE parameter $\leq$0.5 for MOS1 and MOS2, and $\leq$1.4 for the pn, at $10000 <$ PI $< 12000$. These thresholds are slightly higher than usually adopted for EPIC data analysis\footnote{https://www.cosmos.esa.int/web/xmm-newton/sas-thread-epic-filterbackground}, but they enabled us to make good use of the long non-flaring intervals. After filtering, we obtained a good time interval of 32 ks for MOS1 and MOS2, and 26 ks for pn.  We also filtered the event files with the standard conditions ``(FLAG==0) \&\& (PATTERN$<=$4)'' for the pn, and ``(\#XMMEA\_EM \&\& (PATTERN$<=$12)'' for the MOSs ({\it i.e.}, keeping single and double events). 

NGC\,5585 X-1 was detected in both observations. We defined a circular source region of radius 20$^{\prime\prime}$, but with the caveats described in Section 3.1.1, to avoid contamination from a nearby source, and local background regions four times as large as the source region. For each of the two observations, we extracted individual spectra and built associated response and ancillary response files for the pn and MOS cameras with {\it xmmselect}; we then combined the pn and MOS spectra and responses of each observation with {\it epicspeccombine}, to increase the signal-to-noise ratio of possible line features. We grouped the two spectra to a minimum of 25 counts per bin, for subsequent $\chi^2$ spectral fitting with {\sc xspec} \citep{arnaud96} version 12.11.0.



\subsection{{\it HST}}

We observed the field of the candidate ULX bubble in several broad-band and narrow-band filters (Figures 2, 3), with the Wide Field Camera 3 (WFC3), Ultraviolet and VISible light camera (UVIS, chip 2), and Infrared camera (IR). All observations were taken between 2016 April 30 and May 1. We used about a dozen bright, isolated sources in common with the {\it Gaia} Data Release 2 catalog to improve the astrometry of the {\it HST} images. Based on the residual scatter after the re-alignment, we estimate that the {\it HST} coordinates are accurate within $\approx$0$''$.1.

We used the broad-band filters to study the point-like optical counterpart to the ULX (Table 3), and the narrow-band filters to study the line emission from the bubble (Table 4). We retrieved calibrated images (.drc files for WFC3-UVIS, .drz files for WFC3-IR) from the Mikulski Archive for Space Telescopes. We used {\sc ds9} imaging and photometry tools to measure net count rates from the point-like counterpart and the extended bubble. For the counterpart, we used a source extraction radius of 0$''$.16 and a local annular background; we converted the count rates to infinite-aperture values using the online tables of encircled energy fractions. We then converted count rates to magnitudes (in the Vega system) and fluxes, using the latest tables of UVIS zeropoints\footnote{http://www.stsci.edu/hst/wfc3/analysis/uvis\_zpts/uvis1\_infinite} and IR zeropoints\footnote{http://www.stsci.edu/hst/wfc3/ir\_phot\_zpt}.

For the narrow-band filters, we subtracted the continuum as follows. For F502N, we used the F555W image, properly rescaled to account for the different filter width. Likewise, for F164N we used a rescaled F160W image. For F657N and F673N, we used a linear combination of F555W and F814W images, proportional to their respective filter widths. More specifically: i) we selected a region (radius of $\approx$20$^{\prime\prime}$) of the UVIS chip rich in stars but without significant diffuse emission; ii) we measured the count rate from that region in each of the narrow-band filters, and in their corresponding broad-band images; iii) we re-normalized the broad-band images so that their count rates in the test region were equal to the count rates in the associated narrow-band images; iv) we subtracted the re-normalized broad-band images from the narrow-band ones; v) we visually inspected the field around X-1 in the continuum-subtracted narrow-band images to verify that the point-like stars had been properly removed and that any imperfectly-subtracted stellar residuals were negligible (of order of 1 percent) compared with the diffuse nebular emission. We used the {\it fimage} subpackage of the {\sc ftools} software \citep{blackburn95} for these operations, following standard practice.

\begin{figure}
\centering
\includegraphics[height=0.44\textwidth]{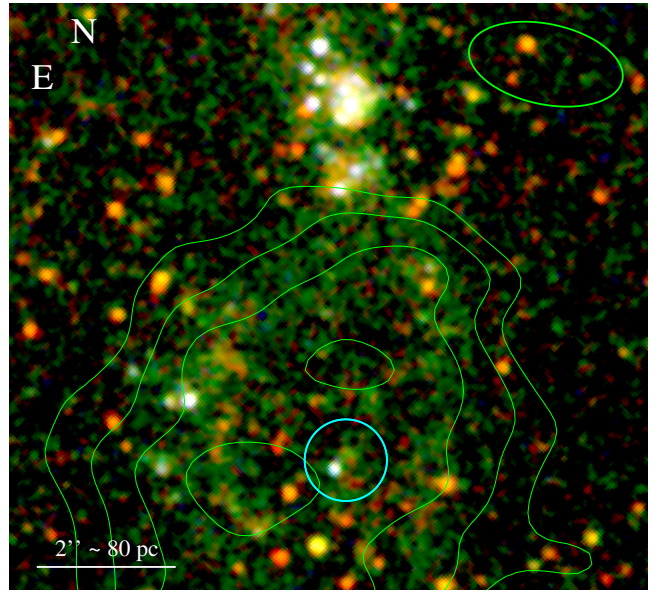}
 \caption{Zoomed-in image of the field around NGC\,5585 X-1, in the WFC3 broad-band filters F814W, F555W and F336W, with radio contours; notice the single blue star-like object inside the 0$''$.6 ULX error radius (cyan circle), which we assume to be the optical counterpart of the X-ray source. Notice also the cluster of young stars immediately above the radio bubble. The green contours and green ellipse were defined in Figure 2.}
  \label{zoom}
\end{figure}

\subsection{LBT}

We observed the nebula with the LBT at Mt Graham, on 2018 June 12--13. 
The LBT consists of twin 8.4-m telescopes, but at the time of our observations only one spectrograph, the Multi-Object Double Spectrograph 1 (MODS1), was in operation. In this detector, a dichroic splits the light beam towards the red and blue spectrograph channels; see \cite{pogge10} for a technical description of the instrument.
We took blue and red spectra with a 0$''$.6 slit (Figure 4), oriented at two different Position Angles (PAs). The first orientation was along the major axis of the bubble (PA $= 355^{\circ}$), including also the young star clusters at the northern tip of the bubble or just outside it. The second orientation was across the bubble (PA $= 65^{\circ}$), in such a way to include the brightest edges of the nebular emission region, roughly 2$''$ to the north-east and the south-west of the ULX. For both orientations, the point-like stellar counterpart of NGC\,5585 X-1 was presumably on the slit, although no trace can be seen in the two-dimensional spectra because of its faintness. 

For the red spectra at both orientations, we used the G670L grating (250 lines/mm), which gives us a resolution of 2300 at 7600 \AA\ (nominal dispersion of 0.845 \AA\ per pixel). For the blue spectra, we used the G400L grating (400 lines/mm), with a resolution of 1850 at 4000 \AA\ (nominal dispersion of 0.51 \AA\ per pixel). At both slit positions, six exposures of 800 s each were obtained.

We bias-subtracted and flat-fielded the raw data with the reduction software  modsCCDRed Version 2.0.1, provided by Ohio State University\footnote{http://www.astronomy.ohio-state.edu/MODS/Software/modsCCDRed/}.  We used the Munich Image Data Analysis System ({\sc midas}: \citealt{warmels92}) for spectral trimming and to determine the wavelength solution. Lines from an Argon lamp yielded the blue channel calibration while the red channel was wavelength calibrated using a mix of Ne, Hg, Kr, Xe and Ar lines. The spectrophotometric standard HZ44 provided an absolute flux calibration common to the blue and red spectra. For further data analysis, such as measurements of line fluxes, line widths and central positions, we used software from both {\sc midas} (in particular, the {\it {integrate/line}} task) and from the Image Reduction and Analysis Facility ({\sc iraf}) Version 2.16 (in particular, the {\it splot} sub-package).

\subsection{VLA}


We observed the field around NGC\,5585 X-1 four times with the VLA between 2015 September 5 and 18 (project ID 15A-142). The phase centre was placed at RA $= 14^{h}19^{m}39^{s}.400$, Dec $= +56^{\circ}41^{\prime}52^{\prime\prime}.70$, slightly offset ($\approx$15$^{\prime\prime}$) from the target. A total integrated time on source of $\approx$3 hr was achieved. For each of the observations, the telescope was in its extended A configuration. Data were taken in the L band, with two contiguous 512-MHz bands observed simultaneously, spanning the 1--2 GHz frequency range. We used 3C\,286 as the bandpass/flux calibrator, while J1400$+$6210 was used as the phase calibrator.

We used the Common Astronomy Software Application \citep[{\sc casa};][]{2007ASPC..376..127M} to perform gain and phase calibration.
All four observations were stacked and imaged using the {\sc clean} algorithm, with Briggs weighting set at a robust value of 1. The final cleaned image (Figure 2, bottom right panel) has a Gaussian restoring beam of 2$^{\prime\prime}.28 \times 1^{\prime\prime}.18$ with a position angle of $77^{\circ}.98$ East of North, and a local rms noise level of $17\,\mu$Jy beam$^{-1}$.

\begin{figure}
\centering
\includegraphics[height=0.44\textwidth]{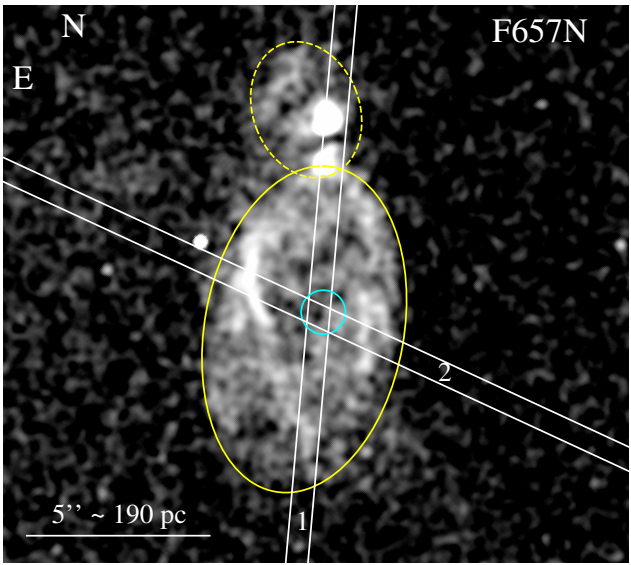}
 \caption{Schematic interpretation of the ionized optical nebula, based on the observed line ratios: the region inside the solid yellow ellipse is the true ULX bubble, dominated by collisional ionization, and with associated synchrotron radio emission; the upper region inside the dashed ellipse is a photo-ionized H II region (ionized by the young stars immediately above the ULX bubble). The positions of the slit in our two LBT observations are overplotted (Obs1 running roughly North-South and Obs2 running roughly East-West). The cyan circle is the location of X-1 (its error radius is 0$''$.6). }
  \label{two_regions}
\end{figure}

\section{Main Results}

\subsection{X-ray properties of NGC\,5585 X-1}

\subsubsection{Identification of X-1 and of a nearby sources}
Our {\it Chandra} and {\it XMM-Newton} study confirmed the presence of a point-like source with an X-ray luminosity of a few times 10$^{39}$ erg s$^{-1}$, located near the centre of the optical bubble. 
We can easily discount the possibility of a chance coincidence with a background X-ray source. At the flux level observed from X-1 ($f_{\rm X} \approx 3 \times 10^{-13}$ erg cm$^{-2}$ s$^{-1}$, as derived later in this Section), we expect $\sim$2--4 background X-ray sources \citep{cappelluti09,luo17} per square degree. This corresponds to a probability of $\sim$1\% to detect one such X-ray source projected behind the whole D25 of NGC\,5585. Then, the probability that such source happened to be randomly projected in the centre of this exceptional optical/radio bubble would be another three orders of magnitude smaller than that.

We then analyzed the possibility of confusion with other sources. From the {\it XMM-Newton}/EPIC-MOS1 and {\it Chandra}/ACIS X-ray contours (Figure 5), there is clearly another (much fainter) source $\approx$18$^{\prime\prime}$ north-east of X-1. It is listed as CXOU J141940.7$+$564150 in the {\it Chandra} Source Catalog \citep{evans10,evans19}, and may be an ordinary high mass X-ray binary in the young stellar population of NGC\,5585. In {\it Chandra}, that source is clearly not a problem for the analysis of X-1. Instead, in {\it XMM-Newton}, it contaminates the emission from X-1 slightly, particularly in EPIC-pn, which has lower spatial resolution than the MOS. We carefully assessed and tried to mitigate the contamination. From our modelling of the {\it Chandra} data, we estimate an average 0.3--10 keV absorbed flux $\approx 1.1 \times 10^{-14}$ erg cm$^{-2}$ s$^{-1}$ for CXOU J141940.7$+$564150 (a factor of 30 fainter than X-1, as we shall see later). In the {\it XMM-Newton} data, the flux was a factor of two higher (although more difficult to estimate, given the stronger source X-1 nearby); both the 3XMM Data Release 7 Serendipitous Source Catalogue from Stacks \citep{traulsen19} and the 3XMM Data Release 8 Serendipitous Source Catalogue \citep{rosen16} list a 0.2--12 keV flux $\approx 2.3$--$2.4 \times 10^{-14}$ erg cm$^{-2}$ s$^{-1}$. [Curiously, the same source is no longer listed in the 4XMM Data Release 9 Serendipitous Source Catalogue \citep{webb20} nor in the 4XMM Data Release 9 Serendipitous Source Catalogue from Stacks \citep{traulsen20}.] For a large X-1 source extraction circle ($\approx$30$^{\prime\prime}$--40$^{\prime\prime}$ radius), CXOU J141940.7$+$564150 would add $\approx$4--7\% of the observed EPIC flux in the two {\it XMM-Newton} observations, which is a significant error and may even affect the observed spectral shape. Thus, we restricted the X-1 extraction radius to 20$^{\prime\prime}$, and we placed a small exclusion circle (10$^{\prime\prime}$ radius) around the position of CXOU J141940.7$+$564150.
With this careful choice of X-1 source region, the estimated flux contamination is reduced to $\approx$1--1.5\%, well below the level of other observational and systematic uncertainties.


\subsubsection{Spectral properties of X-1 from {\it XMM-Newton}}
We start our analysis from the {\it XMM-Newton}/EPIC spectra, which have a higher signal to noise ratio than the {\it Chandra} spectra, and allow a more complex modelling. For every spectral model, we included a photo-electric absorption component ({\it tbabs} in {\sc xspec}, with the abundances of \citealt{wilms00}) fixed at the Galactic line-of-sight value of $2.7 \times 10^{20}$ cm$^{-2}$, and an additional intrinsic {\it tbabs} component left as a free parameter. A simple power-law model is not a good fit (Table 2), because the spectra have significant intrinsic curvature in the EPIC band. A standard disk-blackbody model \citep{shakura73,makishima86} provides a better fit (Figure 6: $\chi^2_{\nu} = 168.5/155$ for June 19 and $\chi^2_{\nu} = 126.7/117$ for June 21), with a characteristic inner disk radius $R_{\rm in} (\cos \theta)^{1/2} \approx (60\pm 5)$ km (June 19) or $R_{\rm in} (\cos \theta)^{1/2} \approx (57\pm 7)$ km (June 21). This is consistent with the innermost stable circular orbit around a stellar-mass black hole (60 km in case of a non-spinning 7-$M_{\odot}$ black hole). 

However, the standard disk solution has a rather high peak temperature ($kT_{\rm in} \approx 1.4$ keV in both observations) and a de-absorbed 0.3--10 keV isotropic luminosity $\approx$2--3 $\times 10^{39}$ erg s$^{-1}$. Taken together, high disk temperature and luminosity formally correspond to super-Eddington mass accretion rates, which are not self-consistent with the standard (sub-Eddington) disk-blackbody model. At such high accretion rates, we expect the disk spectrum to be modified by Comptonization and/or energy advection. Therefore, we generalize the disk model in our spectral fits to take into account those two possibilities. 

First, we approximate a Comptonized disk spectrum with the model {\it simpl} $\times$ {\it diskbb}. The convolution model {\it simpl} \citep{steiner09} includes a fitting parameter for the fraction of seed photons upscattered into a power-law component. The standard disk spectrum corresponds to the case in which the fraction of scattered photons is zero. In our case, the spectrum from the first {\it XMM-Newton} observation has a best-fitting scattering fraction of 0.26, but is barely higher than zero at the 90\% confidence limit (scattering fraction of 0.02). In the (more noisy) second observation, the scattering fraction is consistent with zero at the 90\% confidence level. In both observations, the inner-disk colour temperature $kT_{\rm in} \approx 1.3$ keV and the inner radius $R_{\rm in} (\cos \theta)^{1/2} \approx 70$--80km. We conclude that although the {\it simpl} $\times$ {\it diskbb} model is more physical than the pure {\it diskbb} model, in statistical terms the improvement is marginal at best.

Another model successfully applied to many ULXs \citep{gladstone09} consists of a low-temperature disk component ($kT_{\rm in} \approx 0.1$--0.3 keV) and a warm corona (electron temperature $kT_{e} \approx 1.5$--3 keV). Generally speaking, this model is most suitable when there are two curvature features in the X-ray spectrum: one below 1 keV (associated with a ``soft excess'') and one around 4--6 keV (high-energy roll-over). We tried to apply this type of model to X-1, using {\it diskir} \citep{gierlinski08,gierlinski09} in {\sc xspec}. In both {\it XMM-Newton} spectra, we find no local $\chi^2$ minimum corresponding to this class of cool-disk solutions. The statistically favoured {\it diskir} solution consists again of a dominant disk component with $kT_{\rm in} \approx 1.3$ keV and $R_{\rm in} (\cos \theta)^{1/2} \approx 70$ km, and only a marginal Comptonized component, with a few percent of flux in the high-energy tail. In fact, the Comptonized component is consistent with zero at the 90\% confidence level in the June 21 spectrum. The Comptonized tail is too faint to constrain $kT_{e}$ in either spectrum. Thus, the results of the {\it diskir} model are in perfect agreement with what we found with the {\it simpl} Comptonization model. We will use the {\it diskir} model again later (Section 3.2) to match the X-ray and optical luminosities.




Next, we tried fitting the two spectra with another modified disk model, suitable to super-Eddington accretion: the ''slim disk'' solution
\citep{abramowicz88,kato98,watarai00,watarai01,vierdayanti08}, which includes the effects of advection and radiation trapping. For practical fitting purposes, the slim disk solution is well approximated by the $p$-free disk model ({\it diskpbb} in {\sc xspec}: \citealt{mineshige94,kubota05}), that is a less radiatively efficient disk model in which the temperature scales as $T \propto R^{-p}$, with $p < 0.75$ (the value associated with the radiatively efficient sub-Eddington disk). X-ray spectral studies of Galactic black holes in outburst have shown \citep{kubota04,abe05} that a standard disk transitions to a slim disk when the peak temperature $kT_{\rm in} \gtrsim 1.2$ keV. Both our {\it XMM-Newton} spectra are well-fitted (Table 2) by the $p$-free disk model, with $p \approx 0.7$ and peak colour temperature $kT_{\rm in} \approx 1.5$--1.6 keV. However, the standard solution $p \approx 0.75$ is acceptable within the 90\% confidence limit for both spectra.
We also tried fitting both spectra with the same slim-disk model parameters, locked between the two epochs, except for a  normalization constant; the constant is fixed at 1.0 for the June 19 spectrum, and has a best-fitting value of $0.80\pm0.03$ for the spectrum from June 21 (Table 2). In this case, the best-fitting parameter $p = 0.68^{+0.06}_{-0.04}$, just lower than 0.75 at the 90\% confidence level. We conclude that a slim disk may be preferable to a standard disk for physical reasons, but the statistical difference between the two models is very marginal.

In summary, all three models (standard disk, Comptonized disk and slim disk) give similar de-absorbed luminosities: $L_{\rm X} \approx 3 \times 10^{39}$ erg s$^{-1}$ on June 19, and $L_{\rm X} \approx 2 \times 10^{39}$ erg s$^{-1}$ on June 21, in the 0.3--10 keV band, and peak disk temperatures $\approx$1.3--1.6 keV. Luminosity and peak temperature are self-consistent and typical of a mildly super-Eddington regime, sometimes known as the ``broadened disc'' regime \citep{sutton13}.


\begin{table}
\caption{Summary of our {\it Chandra} and {\it XMM-Newton} observations.}
\vspace{-0.2cm}
\begin{center}
\begin{tabular}{lccc}  
\hline \hline\\[-5pt]    
\multicolumn{4}{c}{{\it Chandra}/ACIS}\\
\hline
ObsID & Obs.~Date &  Exp.~Time  & Net Count Rate \\
 &   &  (ks) & (ct s$^{-1}$, 0.3--8.0 keV)\\
\hline  \\[-5pt]
7150 & 2006 Aug 28 & 5.3 & $(4.7 \pm 0.3) \times 10^{-2}$ \\ [3pt]
19348 & 2017 May 31 & 4.7 &  $(2.8 \pm 0.2) \times 10^{-2}$ \\ [3pt]
\hline\\[-5pt]    
\multicolumn{4}{c}{{\it XMM-Newton}/EPIC}\\
\hline
ObsID & Obs.~Date  &  Exp.~Time  & Net Count Rate \\
 &    & (ks)  & (ct s$^{-1}$, 0.3--10.0 keV)\\
\hline  \\[-5pt]
0762640101 & 2015 Jun 19 & 32.0 (pn) & $(9.7 \pm 0.2) \times 10^{-2}$ \\ [3pt]
           &  & 37.1 (MOS1) & $(2.7 \pm 0.1) \times 10^{-2}$ \\ [3pt]
           &  & 37.1 (MOS2) & $(2.7 \pm 0.1) \times 10^{-2}$ \\ [3pt]
0762640201 & 2015 Jun 21 & 25.5 (pn) & $(7.7 \pm 0.2) \times 10^{-2}$ \\ [3pt]
           &  & 32.4 (MOS1) & $(2.1 \pm 0.1) \times 10^{-2}$ \\ [3pt]
           &  & 32.4 (MOS2) & $(2.1 \pm 0.1) \times 10^{-2}$ \\ [3pt]
\hline\\[-5pt]
\end{tabular} 
\label{tab1}
\end{center}
\end{table}

\begin{figure}
\centering
\includegraphics[width=0.48\textwidth, angle=0]{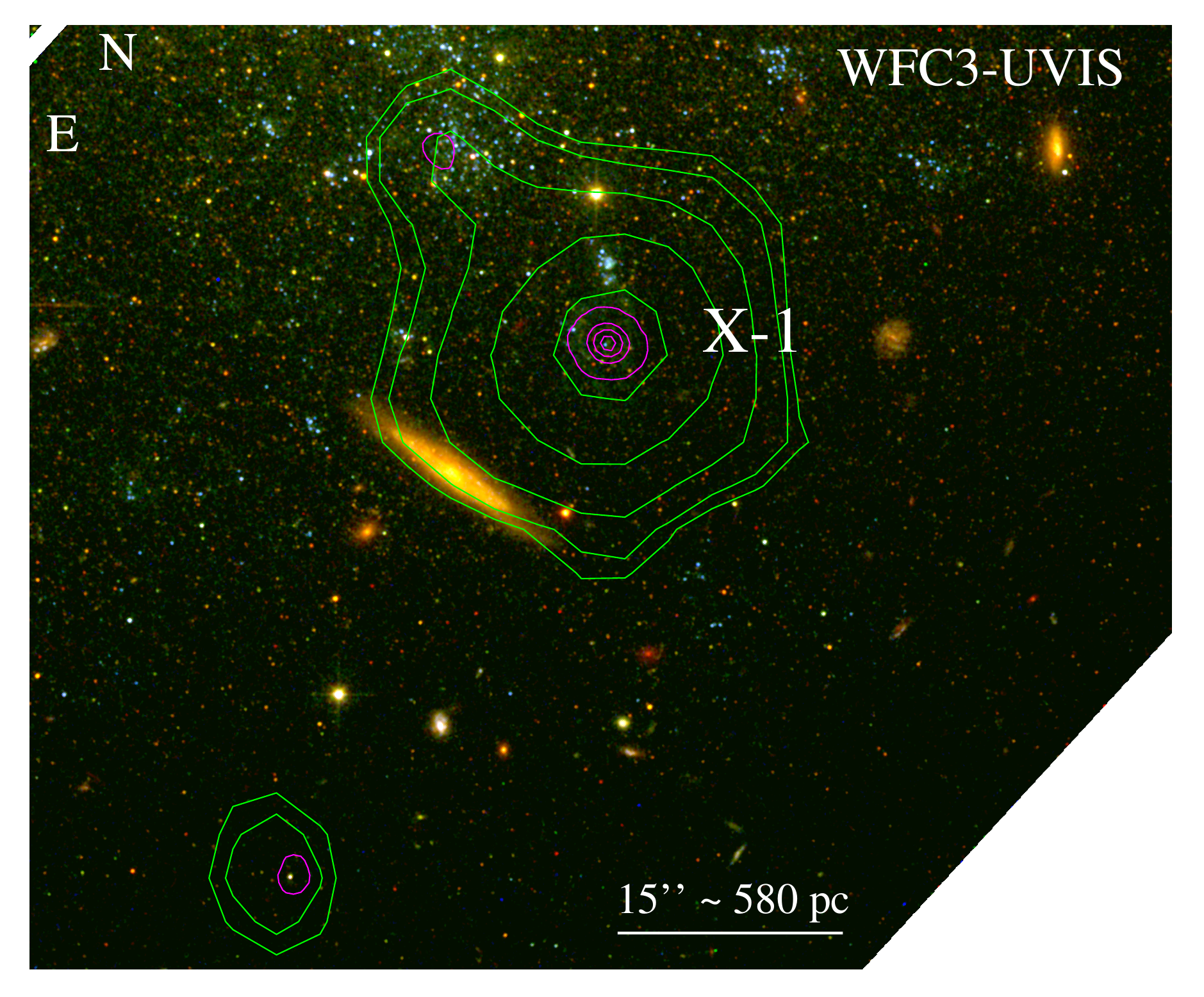}
\caption{X-ray contours of the {\it XMM-Newton}/EPIC-MOS1 (green) and {\it Chandra}/ACIS-S (magenta) count rates in the X-1 region, superposed on an {\it HST}/WFC3 image (F814W = red, F555W = green, F336W = blue). 
{\it XMM-Newton} contours are in log scale, from $10^{-5}$ to $10^{-4}$ MOS1 ct s$^{-1}$ arcsec$^{-2}$. {\it Chandra} contours are in linear scale, from $1.2 \times 10^{-4}$ to $6.4 \times 10^{-3}$ ACIS-S ct s$^{-1}$ arcsec$^{-2}$. The white corner at the bottom right of the image is simply the edge of the WFC3-UVIS chip.
In {\it XMM-Newton} (but not in {\it Chandra}), X-1 is slightly contaminated (at a few percent level) by another source located $\approx$18$^{\prime\prime}$ to the north-east (see Section 3.1.1).
Another faint X-ray source is detected both by {\it XMM-Newton} and by {\it Chandra}, $\approx$40$^{\prime\prime}$ south-east of X-1, but it does not create any contamination issue. Its blue optical counterpart suggests that it is a background quasar. The correspondence between the optical and X-ray position of this background source confirms that the {\it Chandra}/{\it HST} astrometric alignment is better than $\approx$0$^{\prime\prime}$.5 (in turn, the {\it HST} astrometry is aligned to the {\it Gaia} astrometry within $\approx$0$^{\prime\prime}$.1). This supports our identification of the point-like optical counterpart of X-1 (Section 3.2).
}
  \label{xraycontours}
\end{figure}

Estimating the inner radius of a slim disk is less straightforward than for a standard sub-Eddington disk, because it does not coincide with the innermost stable circular orbit \citep{watarai03}. Nonetheless, even for slim disks it is customary to define a ``true'' inner disk radius $R_{\rm in} \approx 3.18 (\kappa/3)^2\, r_{\rm in}$ \citep{vierdayanti08}, where $\kappa$ is the spectral hardening factor, $r_{\rm in}$ is the ``apparent'' radius from the spectral fit, $r_{\rm in} (\cos \theta)^{1/2} = d_{10} \,(N_{\rm dpbb})^{1/2}$, $N_{\rm dpbb}$ is the normalization constant of the {\it diskpbb} model in {\sc xspec}, and $d_{10}$ is the distance in units of 10 kpc. In standard disks, the hardening factor is $\kappa \approx 1.7$ \citep{shimura95}, but for accretion rates at or above the Eddington limit, the hardening factor increases to $\approx$2.5--3 \citep{watarai03,kawaguchi03,shrader03,isobe12}. For a non-rotating black hole, the mass corresponding to a disk radius $R_{\rm in}$ is then $M \approx 1.2 \times R_{\rm in} c^2/(6G) \approx [R_{\rm in}/(7.4~{\rm {km}})]\, M_{\odot}$, where the correction factor 1.2 is also a consequence of the slim disk geometry, more specifically of the fact that a slim disk extends slightly inside the innermost stable orbit \citep{vierdayanti08}. Our viewing angle $\theta$ is unknown, but if we assume that it is not too close to edge-on, the best-fitting value of $R_{\rm in} \approx 120$ km (Table 2) corresponds to a characteristic mass $\approx$15 $M_{\odot}$. Such a system would reach its Eddington limit (and thus show features of the slim disk and/or Comptonized disk regime) when its luminosity reaches $\sim$2 $\times 10^{39}$ erg s$^{-2}$, consistent with what we find in NGC\,5585 X-1.


{\it XMM-Newton} spectra of several other ULXs exhibit features at energies $\sim$0.5--1 keV, consistent with emission and absorption lines from warm outflows, at flux levels of a few percent of the continuum flux \citep{middleton14,middleton15,pinto16,urquhart16,pinto17}.
For NGC\,5585 X-1, we do not find evidence of line residuals in the soft X-ray band, but we cannot draw strong conclusions because of the relatively low number of counts (Table 1). We tried adding an optically thin thermal plasma component ({\it mekal} in {\sc xspec}) to the {\it diskpbb} fit, but it does not improve the fit, and the {\it mekal} normalization is consistent with 0 within the 90\% confidence limit. If we fix the {\it mekal} temperature to $kT = 0.7$ keV, we find a 90\% upper limit of $\approx$6 $\times 10^{37}$ erg s$^{-1}$ for the 0.3--10 keV luminosity of the thermal plasma emission; for a fixed temperature $kT = 1.0$ keV, the 90\% upper limit is $\approx$10$^{38}$ erg s$^{-1}$. We also note that the intrinsic column density is low, $N_{\rm {H,int}} \approx 10^{21}$ cm$^{-2}$. Taken together, the lack of residual line features and the low photo-electric absorption suggest that NGC\,5585 X-1 is not seen through a dense wind (either because of face-on orientation or because the wind is currently weak).

\begin{figure}
\centering
\vspace{-0.7cm}
\includegraphics[height=0.4\textwidth, angle=0]{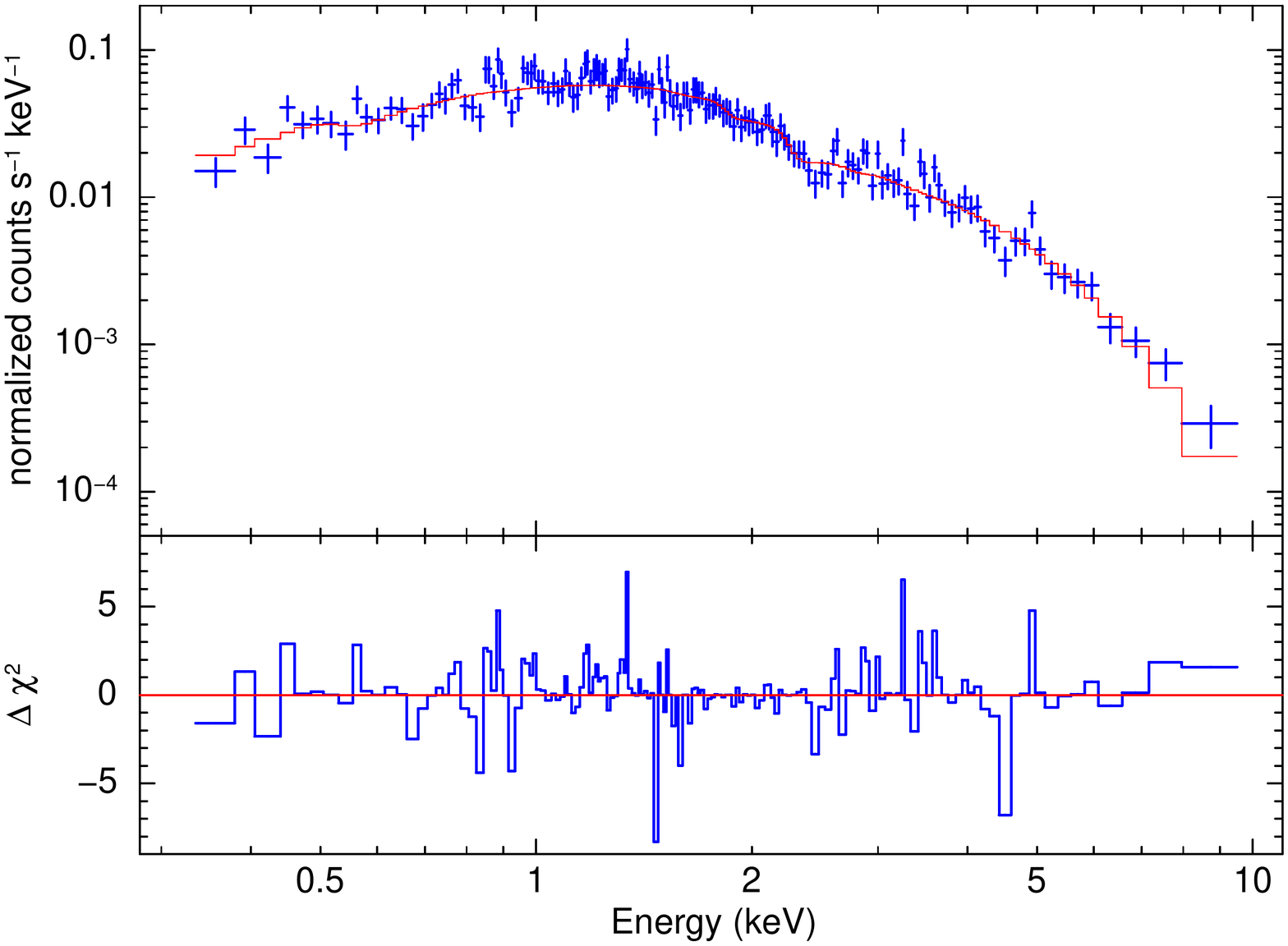}\\[-20pt]
\includegraphics[height=0.4\textwidth, angle=0]{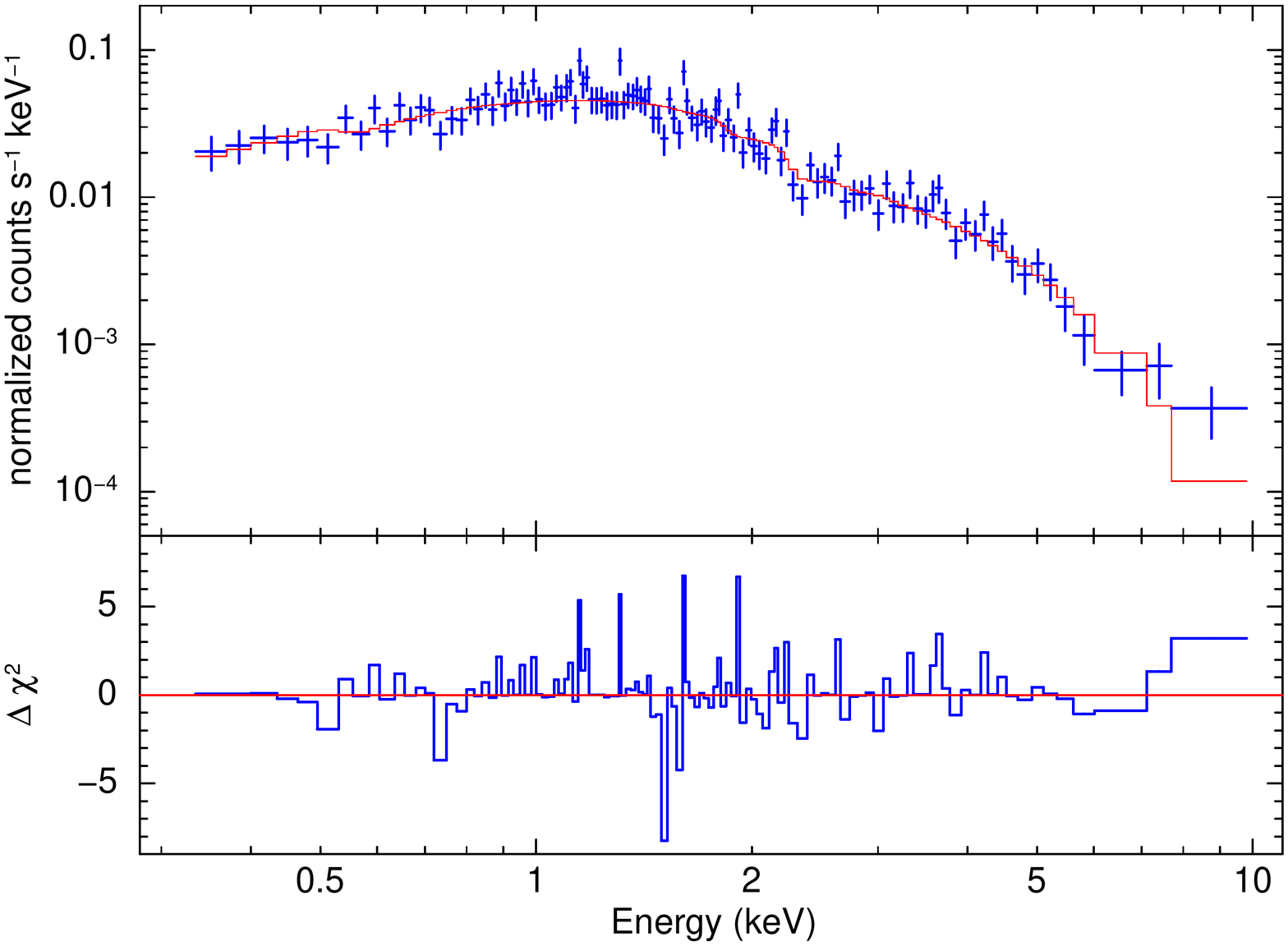}
\vspace{-0.4cm}
 \caption{Top panel: {\it XMM-Newton}/EPIC spectral datapoints and $\chi^2$ residuals, from the 2015 June 19 observation of NGC\,5585 X-1, fitted with a standard disk-blackbody model ({\it diskbb} in {\sc xspec}). The best-fitting inner disk temperature is $T_{\rm in} = (1.44 \pm 0.07)$ keV, and the inferred inner disk radius is $R_{\rm in} (\cos \theta)^{1/2} \approx (60\pm 5)$ km.  Bottom panel: as in the top panel, for the spectrum obtained in the 2015 June 21 observation; here, $kT_{\rm in} = (1.39 \pm 0.09)$ keV, and $R_{\rm in} (\cos \theta)^{1/2} \approx (57\pm 7)$ km. See Table 2 for a more detailed list of spectral parameters, fluxes and luminosities.
 }
  \label{xmmspectra}
\end{figure}

\begin{figure}
\centering
\vspace{-0.7cm}
\includegraphics[height=0.4\textwidth, angle=0]{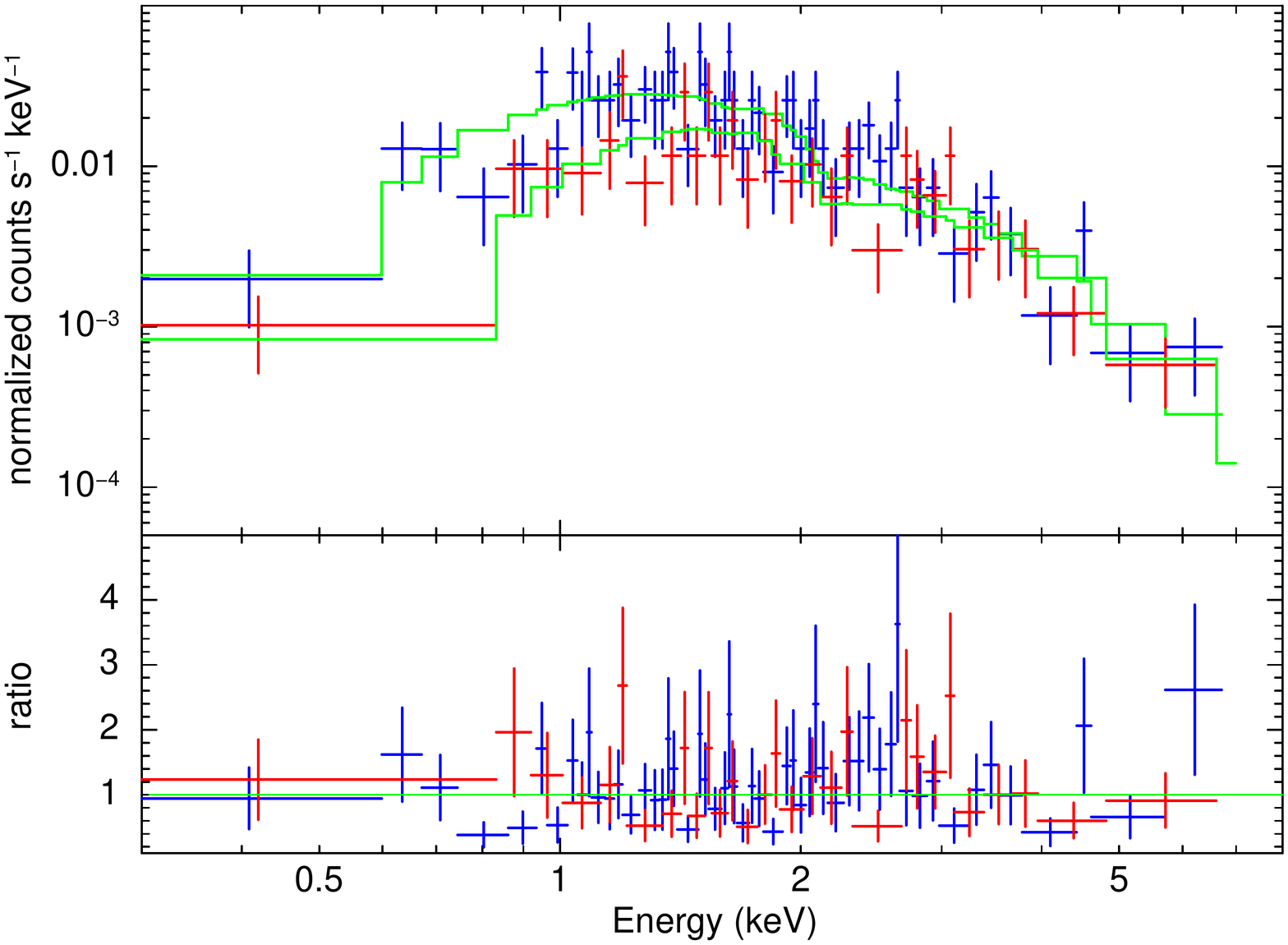}
\vspace{-0.4cm}
 \caption{Simultaneous fit to the {\it Chandra}/ACIS spectra of NGC\,5585 X-1 from 2006 (blue datapoints) and 2017 (red datapoints), with a standard disk-blackbody model, and parameters locked between the two epochs. Here, $kT_{\rm in} = (1.2 \pm 0.2)$ keV, and $R_{\rm in} (\cos \theta)^{1/2} \approx (100\pm 30)$ km.  See Table 3 for the parameter values and luminosity.}
  \label{chandraspectra}
\end{figure}

\subsubsection{Spectral properties of X-1 from {\it Chandra}}

The two {\it Chandra} observations do not have enough counts ($\approx$250 net counts for the 2006 data and $\approx$130 for the 2017 data) for complex, individual spectral modelling. We regrouped each spectrum to 1 count per bin  and fitted it with a power-law model, using the Cash statistics \citep{cash79}. We find that all the parameters (column density, photon index, normalization) are consistent between the two epochs (Table 3); the lower count rate seen in the 2017 observation (Table 1) is only due to the loss of soft X-ray sensitivity in the ACIS detector. Thus, we refitted the two spectra simultaneously with an absorbed {\it diskbb} model (Figure 7),
and locked all parameters between the two epochs, that is we assumed that the emission was identical. This simultaneous slim disk model has a better C statistics (227.9 for 281 degrees of freedom) than the independent power-law models (236.8 for 278 degrees of freedom). The best-fitting peak colour temperature is $kT_{\rm in} \approx (1.2\pm 0.2)$ keV (Table 3). We used the {\it diskbb} model to estimate the observed flux and the de-absorbed luminosity (Table 3); the latter is $L_{\rm X} \approx 3.6 \times 10^{39}$ erg s$^{-1}$, slightly higher than during the 2015 {\it XMM-Newton} observations, but still perfectly consistent with a mildly super-Eddington regime. Irregular flux variability of the amplitude seen between the four {\it XMM-Newton} and {\it Chandra} observations is very common in persistently active ULXs \citep{weng18}.  The inferred inner disk radius is $R_{\rm in} (\cos \theta)^{1/2} \approx (100\pm 30)$ km, consistent with the {\it XMM-Newton} results.

As mentioned earlier (Section 3.1.2), at luminosities $\gtrsim$10$^{39}$ erg s$^{-1}$ and peak disk temperatures $\gtrsim$1.2 keV, theoretical models suggest that the disk emission becomes significantly modified by Comptonization and/or energy advection. As we did with the {\it XMM-Newton} spectrum, we tried convolving the {\it diskbb} spectrum with the {\it simpl} model, but there is no improvement in C statistics, and the scattering fraction is consistent with zero. We also tried the $p$-free model {\it diskpbb}, but again there is no statistical improvement, and the $p$ parameter is consistent with 0.75.


\begin{table}
\caption{Best-fitting parameters of the {\it XMM-Newton}/EPIC spectra of X-1. 
The Galactic absorption is fixed at $N_{\rm {H,Gal}} = 2.7 \times 10^{20}$ cm$^{-2}$.} 
\vspace{-0.3cm}
\begin{center}  
\begin{tabular}{lcc} 
 \hline 
\hline \\[-8pt]
  Model Parameters      &      \multicolumn{2}{c}{Values} \\
  & 2015 June 19       &       2015 June 21       \\
\hline\\[-9pt]
\multicolumn{3}{c}{{\it tbabs} $\times$ {\it tbabs} $\times$ {\it po}}\\
\hline\\[-6pt]
   $N_{\rm {H,int}}$   ($10^{22}$ cm$^{-2}$)   &  $ 0.29^{+0.03}_{-0.03}$ &     $ 0.24^{+0.04}_{-0.04}$ \\[4pt]
   $\Gamma$      &  $ 1.88^{+0.07}_{-0.06}$       &   $ 1.85^{+0.09}_{-0.09}$   \\[4pt] 
   $N_{\rm {po}}^a $ 
              &  $ 8.5^{+0.6}_{-0.6}$   &   $ 6.3^{+0.6}_{-0.6}$ \\[4pt]
   $\chi^2$/dof     &      $234.7/155$ (1.51)            &       $165.0/117$ (1.41) \\
   \hline\\[-9pt]
   \multicolumn{3}{c}{{\it tbabs} $\times$ {\it tbabs} $\times$ {\it simpl} $\times$ {\it diskbb}}\\
\hline\\[-6pt]
   $N_{\rm {H,int}}$   ($10^{22}$ cm$^{-2}$)   &  $ 0.07^{+0.02}_{-0.02}$ &     $ 0.04^{+0.03}_{-0.02}$ \\[4pt]
   $\Gamma$      &  $ 1.1^{+5.0}_{-0.1}$       &   [1.1]   \\[4pt] 
   FracScatt    &   $ 0.26^{+0.08}_{-0.24}$            & $<0.41$  \\[4pt]
   $kT_{\rm {in}} $ (keV)
              &  $ 1.29^{+0.13}_{-0.31}$   &   $ 1.30^{+0.17}_{-0.18}$ \\[4pt]
   $N_{\rm {dbb}}$  ($10^{-3}$ km$^2$)$^b$
              &  $ 7.4^{+3.0}_{-3.1}$   &   $ 5.1^{+5.4}_{-2.2}$ \\[4pt]
   $R_{\rm {in}}\sqrt{\cos \theta} $  (km)$^c$ &  $ 82^{+15}_{-19}$   &   $ 68^{+30}_{-17}$ \\[4pt]
   $\chi^2$/dof     &      $165.2/153$ (1.08)            &       $126.2/116$ (1.09) \\
   \hline\\[-7pt]
\multicolumn{3}{c}{{\it tbabs} $\times$ {\it tbabs} $\times$ {\it diskpbb}}\\
\hline\\[-5pt]
   $N_{\rm {H,int}}$   ($10^{22}$ cm$^{-2}$)   &  $ 0.11^{+0.02}_{-0.05}$ &     $ 0.06^{+0.06}_{-0.06}$ \\[4pt]
   $kT_{\rm in}$ (keV)     &  $ 1.58^{+0.21}_{-0.16}$       &   $ 1.46^{+0.27}_{-0.19}$   \\[4pt] 
   $p$    &   $ 0.68^{+0.07}_{-0.06}$            & $ 0.71^{+0.15}_{-0.09}$  \\[4pt]
   $N_{\rm {dpbb}} $  ($10^{-3}$ km$^2$)$^b$ &  $ 2.2^{+2.2}_{-1.1}$   &   $ 2.6^{+4.0}_{-1.6}$ \\[4pt]
   $R_{\rm {in}}\sqrt{\cos \theta} $  (km)$^d$ &  $ 119^{+48}_{-35}$   &   $ 128^{+79}_{-50}$ \\[4pt]
   $\chi^2/$dof     &      $166.2/154$ (1.08)            &       $126.4/116$ (1.09) \\[4pt]
   $f_{0.3-10}$ ($10 ^{-13}$ erg cm$^{-2}$ s$^{-1}$)$^e$ & $ 3.38^{+0.13}_{-0.14}$ & $ 2.57^{+0.14}_{-0.14}$ \\[4pt]
   $L_{0.3-10}$ ($10 ^{39}$ erg s$^{-1}$)$^f$ & $ 2.91^{+0.19}_{-0.18}$ & $ 2.13^{+0.20}_{-0.18}$  \\[4pt]
      \hline\\[-9pt]
\multicolumn{3}{c}{{\it tbabs} $\times$ {\it tbabs} $\times$ {\it const} $\times$ {\it diskpbb}}\\
\hline\\[-6pt]
   $N_{\rm {H,int}}$   ($10^{22}$ cm$^{-2}$)   &  \multicolumn{2}{c}{$ 0.09^{+0.04}_{-0.04}$} \\[4pt]
   $kT_{\rm in}$ (keV)     &  \multicolumn{2}{c}{$ 1.55^{+0.15}_{-0.13}$}      \\[4pt] 
   $p$    &   \multicolumn{2}{c}{$ 0.68^{+0.06}_{-0.04}$}   \\[4pt]
   $C$   & [1.00]   &   $ 0.80^{+0.03}_{-0.03}$ \\[4pt]
   $N_{\rm {dpbb}} $  ($10^{-3}$ km$^2$)$^b$ &  \multicolumn{2}{c}{$ 2.3^{+1.8}_{-1.0}$ }  \\[4pt]
   $R_{\rm {in}}\sqrt{\cos \theta} $  (km)$^d$ &    \multicolumn{2}{c}{$ 123^{+40}_{-31}$} \\[4pt]
   $\chi^2/$dof     &       \multicolumn{2}{c}{$299.7/273$ (1.10)}  \\[4pt]
   $f_{0.3-10}$ ($10 ^{-13}$ erg cm$^{-2}$ s$^{-1}$)$^e$ & $ 3.33^{+0.12}_{-0.12}$ & $ 2.65^{+0.10}_{-0.10}$ \\[4pt]
   $L_{0.3-10}$ ($10 ^{39}$ erg s$^{-1}$)$^f$ & $ 2.84^{+0.16}_{-0.14}$ & $ 2.26^{+0.13}_{-0.11}$  \\[4pt]
\hline 
\vspace{-0.5cm}
\end{tabular}
\end{center}
\begin{flushleft} 
$^a$: units of $10^{-5}$ photons keV$^{-1}$ cm$^{-2}$ s$^{-1}$ at 1 keV.\\
$^b$: $N_{\rm {dbb}} = (r_{\rm{in}}/d_{10})^2 \cos \theta$, where $r_{\rm {in}}$ is the apparent inner disk radius in km, $d_{10}$ the distance to the source in units of 10 kpc (here, $d_{10} = 800$), and $\theta$ is our viewing angle ($\theta = 0$ is face-on). $N_{\rm {dpbb}}$ is defined the same way.\\
$^c$: $R_{\rm {in}} \approx 1.19 r_{\rm in}$ for a standard disk \citep{kubota98}.\\ 
$^d$: $R_{\rm {in}} \approx 3.18 \, (\kappa/3)^2 \, r_{\rm in}$ for a slim disk \citep{vierdayanti08,isobe12}. Here we assume $\kappa = 3$.\\ 
$^e$: observed fluxes in the 0.3--10 keV band\\
$^f$: isotropic de-absorbed luminosities in the 0.3--10 keV band, defined as $4\pi d^2$ times the de-absorbed fluxes.
\end{flushleft}
\end{table}





\begin{table}
\caption{Best-fitting parameters of the {\it Chandra}/ACIS spectra of X-1. As in Table 2, the Galactic absorption is fixed at $N_{\rm {H,Gal}} = 2.7 \times 10^{20}$ cm$^{-2}$.} 
\vspace{-0.3cm}
\begin{center}  
\begin{tabular}{lcc} 
 \hline 
\hline \\[-8pt]
  Model Parameters      &      \multicolumn{2}{c}{Values} \\
  & 2006 August 28       &       2017 May 31       \\
\hline\\[-9pt]
\multicolumn{3}{c}{{\it tbabs} $\times$ {\it tbabs} $\times$ {\it po}}\\
\hline\\[-9pt]
   $N_{\rm {H,int}}$   ($10^{22}$ cm$^{-2}$)   &  $ 0.50^{+0.21}_{-0.18}$ &     $ 0.48^{+0.47}_{-0.39}$ \\[4pt]
   $\Gamma$      &  $ 1.91^{+0.36}_{-0.34}$       &   $ 2.15^{+0.54}_{-0.49}$   \\[4pt] 
   $N_{\rm {po}}^a $ 
              &  $ 1.3^{+0.6}_{-0.4}$   &   $ 1.5^{+1.2}_{-0.6}$ \\[4pt]
   C-stat/dof     &      $142.3/153$ (0.93)            &       $94.6/125$ (0.76) \\
   \hline\\[-9pt]
\multicolumn{3}{c}{{\it tbabs} $\times$ {\it tbabs} $\times$ {\it diskbb}}\\
\hline\\[-6pt]
$N_{\rm {H,int}}$   ($10^{22}$ cm$^{-2}$)   &  \multicolumn{2}{c}{$ 0.25^{+0.12}_{-0.11}$}  \\[4pt]
   $kT_{\rm in}$ (keV)     &  \multicolumn{2}{c}{$ 1.18^{+0.20}_{-0.15}$}\\[4pt] 
   $N_{\rm {dbb}} $  ($10^{-3}$ km$^2$)$^b$ &  \multicolumn{2}{c}{$ 11.8^{+8.9}_{-5.4}$}\\[4pt]
   $R_{\rm {in}}\sqrt{\cos \theta} $  (km)$^c$ &  \multicolumn{2}{c}{$103^{+34}_{-26}$}\\[4pt]
   C-stat/dof     &      \multicolumn{2}{c}{$227.9/281$ (0.81)}\\[4pt]
   $f_{0.3-10}$ ($10 ^{-13}$ erg cm$^{-2}$ s$^{-1}$)$^d$ 
       & \multicolumn{2}{c}{$ 3.9^{+0.5}_{-0.5}$}\\[4pt]
   $L_{0.3-10}$ ($10 ^{39}$ erg  s$^{-1}$)$^e$ 
      & \multicolumn{2}{c}{$ 3.6^{+0.5}_{-0.3}$}\\[2pt]
\hline 
\vspace{-0.5cm}
\end{tabular}
\end{center}
\begin{flushleft} 
$^a$: units of $10^{-4}$ photons keV$^{-1}$ cm$^{-2}$ s$^{-1}$ at 1 keV.\\
$^b$: $N_{\rm {dbb}} = (r_{\rm{in}}/d_{10})^2 \cos \theta$, where $r_{\rm{in}}$ is the apparent inner disk radius in km, $d_{10}$ the distance to the source in units of 10 kpc (here, $d_{10} = 800$), and $\theta$ is our viewing angle ($\theta = 0$ is face-on).\\
$^c$: $R_{\rm {in}} \approx 1.19 r_{\rm in}$ for a standard disk \citep{kubota98}.\\
$^d$: observed fluxes in the 0.3--10 keV band\\
$^e$: isotropic de-absorbed luminosities in the 0.3--10 keV band, defined as $4\pi d^2$ times the de-absorbed fluxes.
\end{flushleft}
\end{table}

\subsection{Point-like optical counterpart}

The short exposure times and the moderately off-axis location of NGC\,5585 X-1 in both {\it Chandra} observations are the main reasons why we cannot substantially improve the astrometric solution of the X-ray images using optical/X-ray associations. Thus, we took the default astrometry of the ACIS event files, and determined the centroid of X-1 in both observations. We obtain a position R.A.(J2000) $= 14^h 19^m 39^s.38$, Dec.(J2000) $=56^{\circ}41^{\prime}37^{\prime\prime}.7$ (the same in both observations), with a 90\% error radius of $0^{\prime\prime}.6$. (The 90\% uncertainty radius of an individual {\it Chandra} observation is $\approx$0$^{\prime\prime}.8$\footnote{https://cxc.harvard.edu/cal/ASPECT/celmon/}, but here we are taking the average of two observations.)
There is only one bright, blue optical source (Figure 3) that stands out inside the X-ray error circle, in the {\it HST} images: it is located  at R.A.(J2000) $= 14^h 19^m 39^s.41$, Dec.(J2000) $=56^{\circ}41^{\prime}37^{\prime\prime}.6$ (error radius of $0^{\prime\prime}.1$). 

We looked for other {\it Chandra} sources with an obvious point-like {\it HST} counterpart, which could help us refine the X-1 astrometry, from their relative offsets. There is an X-ray source (listed as CXOGSG J141941.9$+$564101 in \citealt{wang16}) coincident with an obvious, bright point-like {\it HST} counterpart (a background quasar), about 40$^{\prime\prime}$ south-east of X-1 (Figure 5); however, it only has 13 ACIS counts scattered within a 2$^{\prime\prime}$ radius, and we cannot confidently fit its central position to better than about 0$^{\prime\prime}$.5. There is another faint {\it Chandra} source about 82$^{\prime\prime}$ north-west of X-1, clearly associated with another likely quasar in the {\it HST} images, at R.A.(J2000) $= 14^h 19^m 30^s.51$, Dec.(J2000) $=56^{\circ}42^{\prime}13^{\prime\prime}.1$; but it has only 6 ACIS counts, so again we cannot determine the X-ray position with great accuracy. Nonetheless, the coincidence between those two other faint sources and their {\it HST} counterparts within about 0$^{\prime\prime}$.5 confirms the good astrometric alignment, supports our identification of the optical counterpart of X-1, and suggests that our error circle of about 0$^{\prime\prime}$.6 is a safe, conservative estimate.

The optical brightness and colour of the optical counterpart of X-1 are consistent with those of other ``blue'' ULX counterparts \citep{tao11,gladstone13}. Correcting only for foreground Galactic extinction, we infer absolute brightnesses $M_{F336W} = -5.79 \pm 0.04$ mag, $M_{F555W} = -4.29 \pm 0.03$ mag, $M_{F814W} = -3.65 \pm 0.10$ mag (Table 4). The X-ray to optical flux ratio (defined as in \citealt{maccacaro82}) is $\log(f_{\rm X}/f_V) \approx 2.80$, which is also typical of ULX counterparts \citep{tao11,gladstone13}. Of course we are aware that the X-ray to optical flux ratio was not measured simultaneously, and we have no information on how the optical luminosity may vary in response to the observed small changes in X-ray luminosity; but we only need an order-of-magnitude estimate of this ratio to conclude that X-1 is not for example a background AGN (which have typical ratios $-1 \lesssim \log(f_{\rm X}/f_V) \lesssim 1$).

As is usually the case for blue ULX counterparts, it is hard to tell (especially with only three photometric datapoints) whether the optical source is a young, massive star, or the irradiation-dominated outer accretion disk (or a combination of both). First, we assumed that the emission is entirely from a single donor star. We used the stellar isochrones from the Padova group \citep{marigo17}, available online\footnote{http://stev.oapd.inaf.it/cgi-bin/cmd}; we assumed a subsolar metal abundance ($Z = 0.008$), expected for the outskirts of a small disk galaxy, as suggested by the analysis of \cite{ganda07} and also confirmed by our subsequent spectral analysis (Section 3.3.7). We found that the optical magnitudes are consistent either with an O8.5-O9 main sequence star, or with a B0 subgiant, with a characteristic temperature $T \approx (30,000 \pm 2,000)$ K. The mass range goes from $M \approx 25 M_{\odot}$ for an O8.5 V star (age $\lesssim 3$ Myr), to $M \approx 15 M_{\odot}$ for a B0IV star (age $\approx 12$ Myr). Alternatively, we used the {\it diskir} model in {\sc xspec}, fitted to both the X-ray and optical datapoints. The two model parameters that constrain the optical/UV flux are the outer disk radius (expressed as the ratio of outer/inner disk radii) and the reprocessing fraction, which tells us how much of the illuminating X-ray flux is intercepted and re-emitted by the disk (mostly in the optical/UV band). We found that an accretion disk with an outer radius of $\sim$a few $10^{11}$--$10^{12}$ cm, and an X-ray reprocessing fraction $f \sim 10^{-2}$ also produces optical emission consistent with the observed colours; such a high reprocessing factor is not unusual in ULXs \citep{sutton14}, possibly because outflows enhance the fraction of X-ray photons down-scattered onto the surface of the outer disk. Thus, we cannot distinguish between the star and irradiated disk scenarios. One (weak) argument in favour of a significant disc contribution to the optical counterpart is that this is the only blue optical source within $\approx$100 pc around X-1; massive stars rarely come in isolation.

\begin{table}
\caption{Optical brightness of the point-like counterpart of NGC\,5585 X-1, determined from the {\it HST} observations.}
\vspace{-0.4cm}
\begin{center}
\begin{tabular}{lccc}  
\hline \hline\\[-5pt]    
Filter &  Exp.~Time &   Apparent Brightness  & Absolute Brightness$^a$\\
 &  (s)  & (mag) & (mag)\\
\hline  \\[-5pt]
F336W & 1200 &  $23.80 \pm 0.04$ & $-5.79 \pm 0.04$ \\ [3pt]
F555W & 1245  &  $25.27 \pm 0.03$ & $-4.29 \pm 0.03$ \\ [3pt] 
F814W & 1350  &  $25.90 \pm 0.10$ & $-3.65 \pm 0.10$ \\ [3pt]
\hline
\vspace{-0.5cm}
\end{tabular} 
\end{center}
\begin{flushleft} 
\footnotesize{
$^a$: corrected for Galactic foreground reddening $E(B-V) = 0.014$ mag.}
\end{flushleft}
\end{table}

\subsection{ULX optical bubble versus H II region}

\subsubsection{Morphology}
From the {\it HST} image of the bubble in the F657N band (Figures 2, 4), we estimate a diameter of $\approx 9^{\prime\prime}.0$ in the north-south direction and $\approx 5^{\prime\prime}.7$ in the east-west direction. X-1 is roughly in the centre of the bubble, $\approx 4^{\prime\prime}$ from the northern edge along the major axis, and $\approx 3^{\prime\prime}$ from the eastern edge along the minor axis. At the assumed distance of 8 Mpc, the bubble size corresponds to $\approx$350 $\times$ $\approx$220 pc, and its volume is $\approx$3 $\times 10^{62}$ cm$^3$ for a prolate spheroid approximation ({\it i.e.}, with radii of 110, 110 and 175 pc). An essentially identical size ($\approx$350 $\times 230$ pc) is independently estimated from the 1.5-GHz VLA image (Figure 2). It is one of the largest ULX bubbles known to-date, in the same class as the ``classical" bubbles around NGC\,1313 X-2, Holmberg IX X-1 and NGC\,7793-S26. This size does not include the additional, smaller photo-ionized nebula at the northern end, energized by a small cluster of young stars.

The narrow-band {\it HST} imaging already visually suggests that the ULX bubble is shock-ionized, from its relatively high brightness in [S {\footnotesize{II}}]$\lambda6716,6731$ (F673N filter) and [Fe {\footnotesize{II}}]$\lambda$1.64$\mu$m (F164N) (Figure 2); instead, there is no discernible emission from those two lines in the northern photo-ionized region. Radio emission is also associated only with the shock-ionized ULX bubble.

In the rest of this Section 3.3, we will use the LBT spectra (Figures 8,9) to calculate intensity ratios between the main diagnostic lines, and determine quantities such as temperature, density and metal abundance of the emitting gas. We will use the {\it HST} imaging data to measure the total flux and luminosity in the main lines.




\begin{figure}
\centering
\includegraphics[height=0.47\textwidth, angle=270]{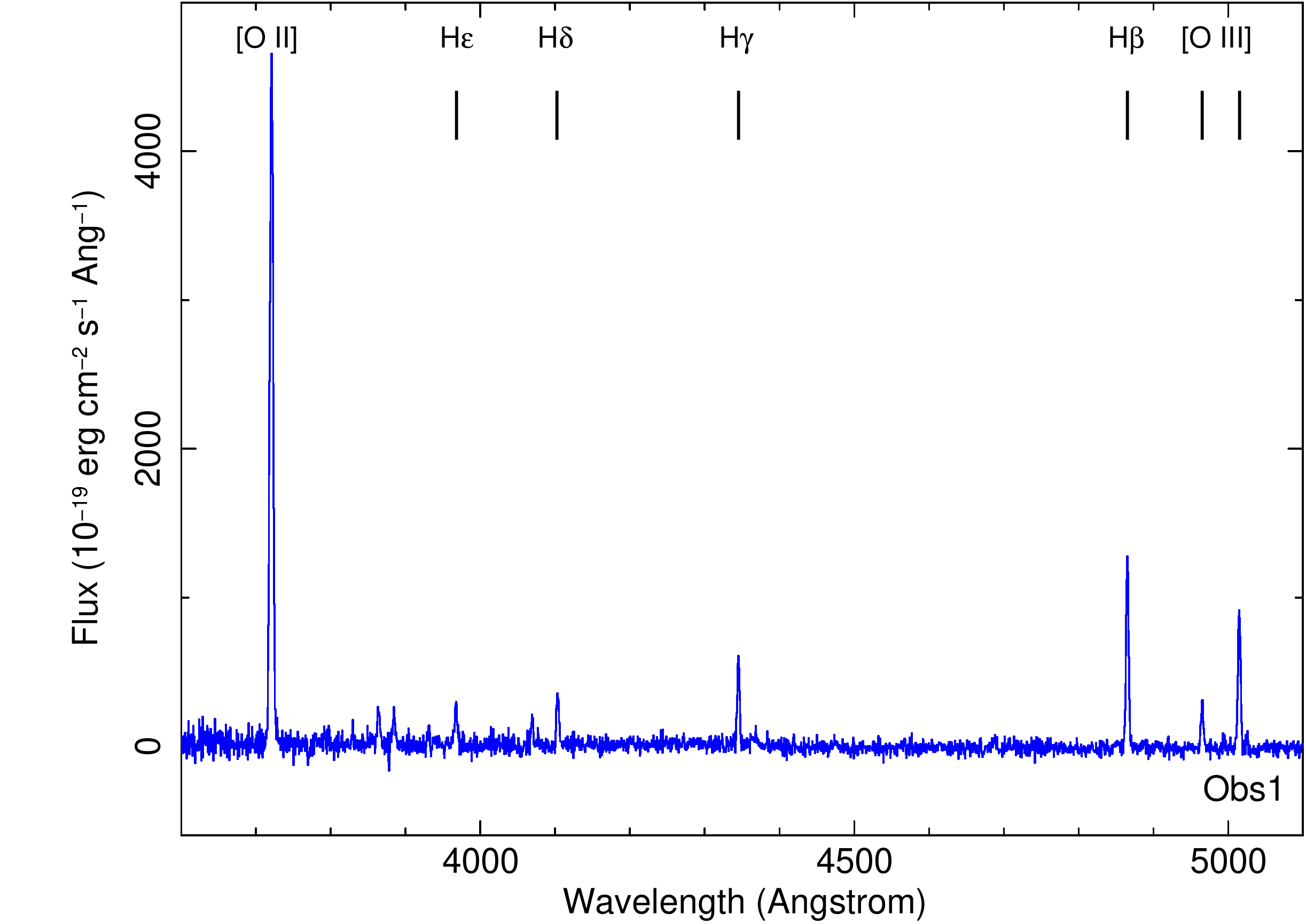}\\[5pt]
\includegraphics[height=0.47\textwidth, angle=270]{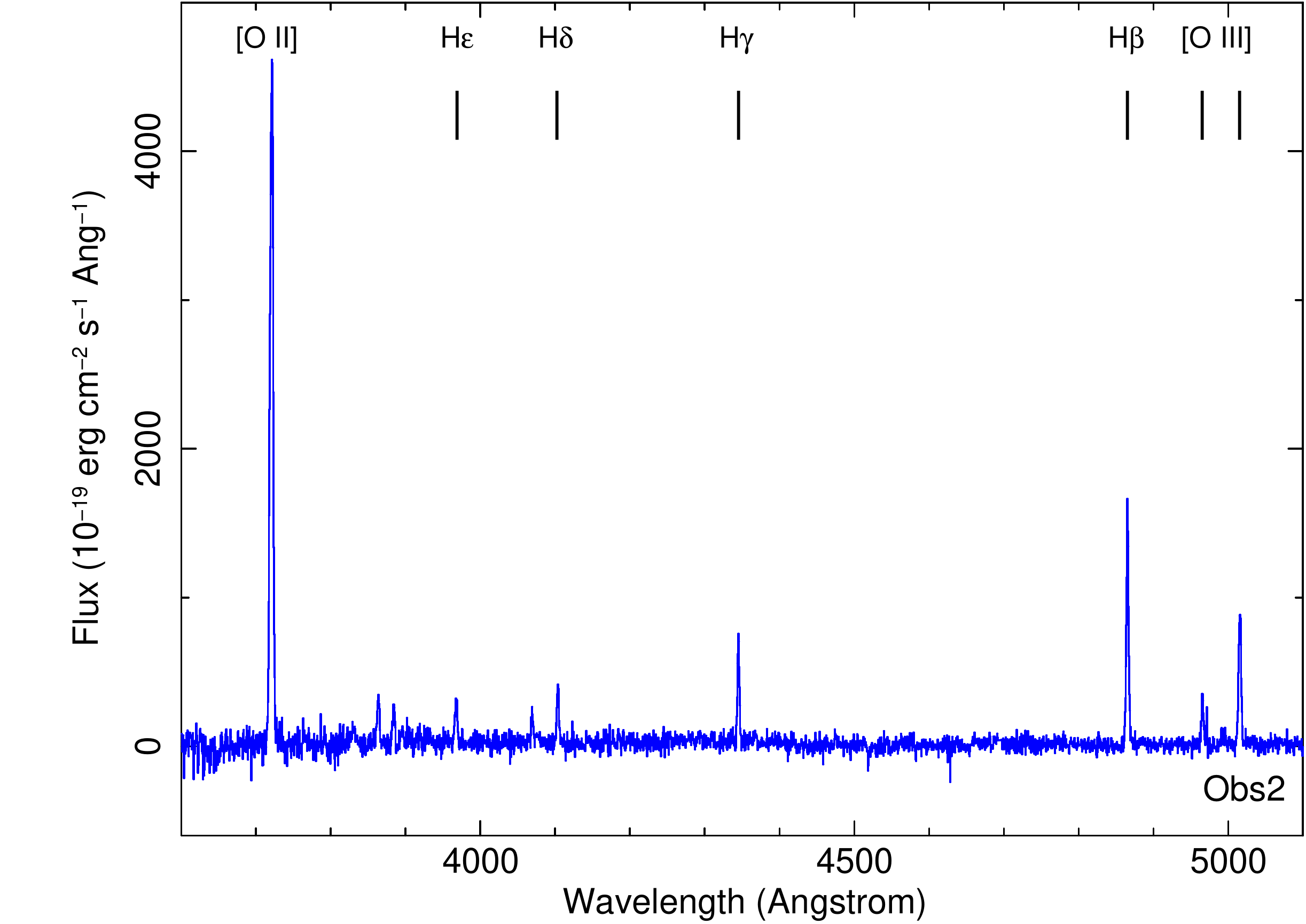}
 \caption{Top panel: blue portion of the LBT spectrum of the ULX bubble (excluding the H II region) with the slit oriented along the major axis of the bubble (Obs1). Bottom panel: blue spectrum with the slit oriented along the minor axis (Obs2). In both panels, the main lines are identified.}
  \label{lbtblue}
\end{figure}

\begin{figure}
\centering
\includegraphics[height=0.47\textwidth, angle=270]{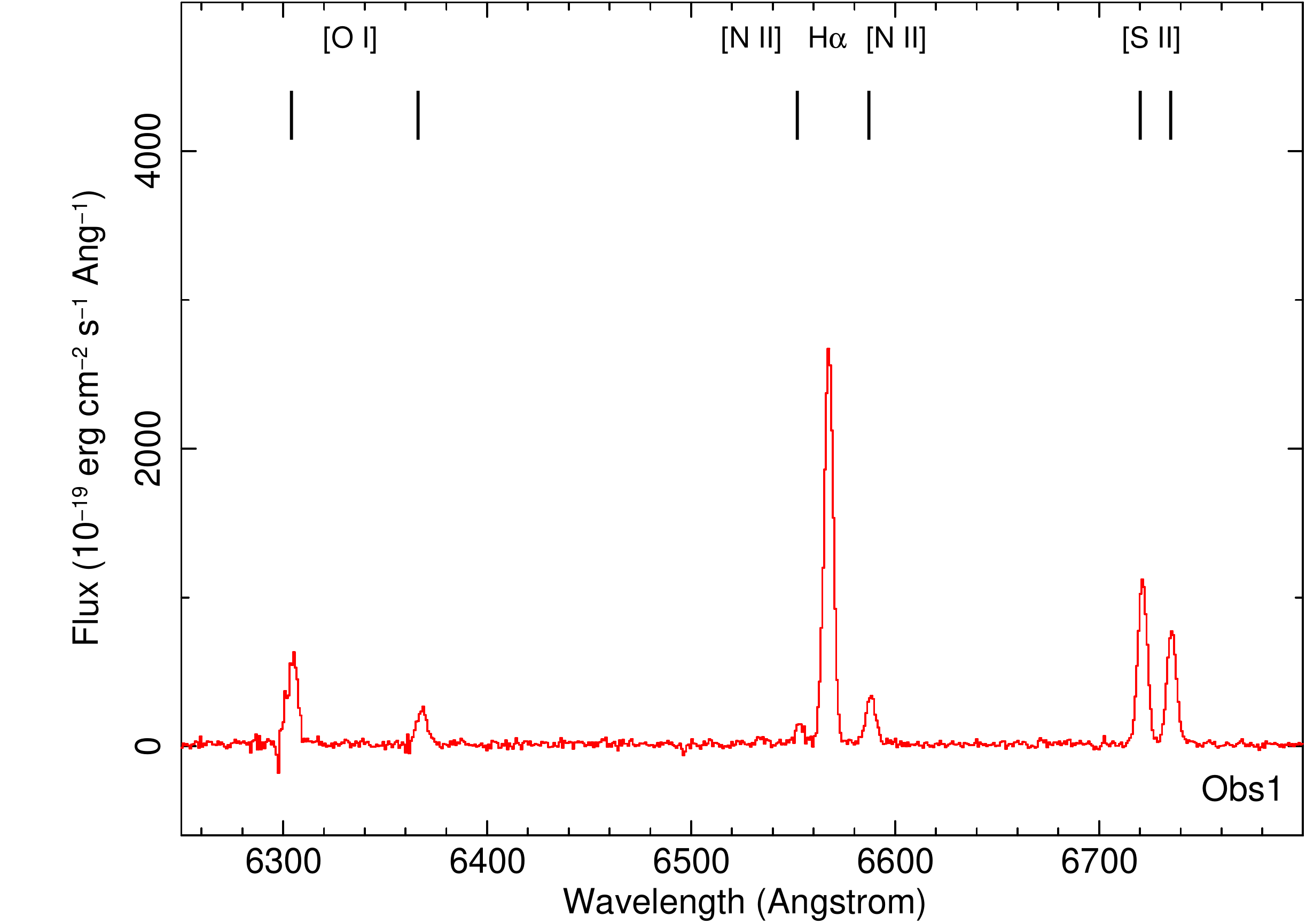}\\[5pt]
\includegraphics[height=0.47\textwidth, angle=270]{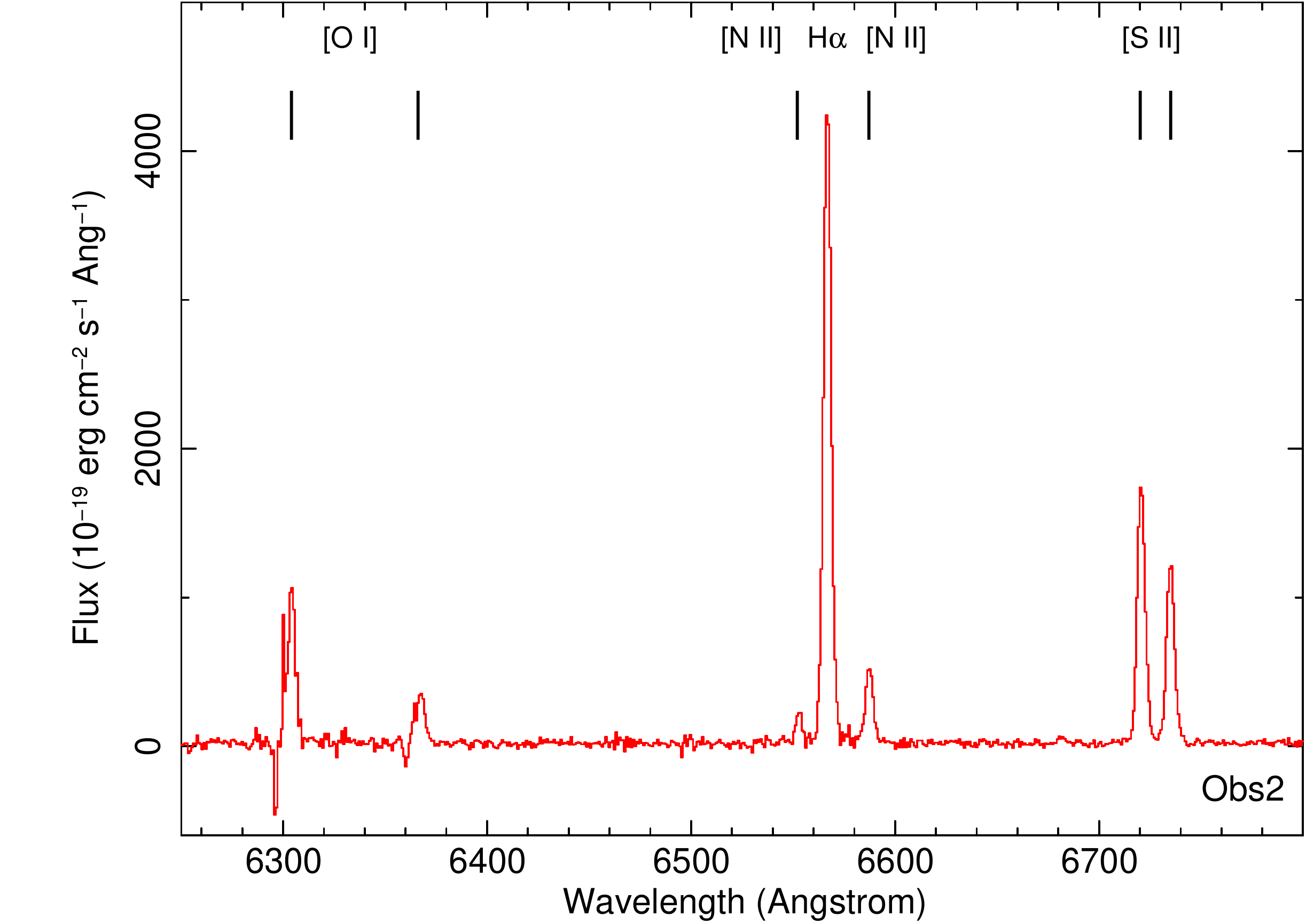}
 \caption{Top panel: red portion of the LBT spectrum with the slit oriented along the major axis of the bubble (Obs1), excluding the H II region. Bottom panel: red spectrum with the slit oriented along the minor axis (Obs2). In both panels, the main lines are identified.}
  \label{lbtred}
\end{figure}

\subsubsection{Diagnostic line ratios}
The first basic classification of an optical nebula is based on the ratio between the flux in the sulphur doublet [S {\footnotesize{II}}]$\lambda\lambda$6716,6731 and in H$\alpha$ \citep{mathewson73,blair97}. High values of this ratio ($\gtrsim$0.3) imply collisional ionization. 
Because of the small wavelength difference between the lines, the differential reddening is negligible and the ratio is well estimated directly from the observed flux values.  We measured the flux ratio along the slit in the both the Obs1 (major axis of the nebula, roughly north to south) and Obs2 (minor axis, roughly east to west) positions. For Obs1, we obtained a [S {\footnotesize{II}}]/H$\alpha$ ratio of $\approx$0.72 (defining the shock-ionized ULX bubble) between $\approx$4$^{\prime\prime}.5$ to the south and $\approx$3$^{\prime\prime}$ to the north of the ULX position, measured along the slit (Table 5 and top panel of Figure 10). Further to the north direction, the flux ratio drops rapidly, reaching an average value of $\approx$0.14 for slit positions between $\approx$4$^{\prime\prime}.5$ and $6^{\prime\prime}.2$ from the ULX; this is the location of the photo-ionized H {\footnotesize{II}} region. In the Obs2 position, the slit intercepts only the ULX bubble; we measure an average flux ratio of $\approx$0.71 (Table 5).

The second important line ratio considered in our analysis is [S {\footnotesize{II}}]$\lambda 6716/6731$, an indicator of electron density \citep{osterbrock06}. We measured (Table 5) a ratio of $\approx$1.47 in the bubble region of Obs1, $\approx$1.44 for Obs2, and $\approx$1.5 for the H {\footnotesize{II}} region of Obs1. All three values fall in the asymptotic low-density regime, $n_e \lesssim$ a few $\times 10$ cm$^{-3}$. 
We will show later (Section 4.1) that the density of the ambient (non-shocked) interstellar medium is $\sim$1 cm$^{-3}$. Standard magnetohydrodynamic relations show that for a radiative shock in the presence of magnetic fields, the compression factor (density ratio) across the shock is $\rho_2/\rho_1 \approx b \mathscr{M}^2$, where $\mathscr{M}$ is the upstream isothermal Mach number; the exact value of the proportionality constant $b$ depends on the small-scale properties of turbulence and magnetic field, but it is always $b < 1$ ($b \approx 1$ in the non-magnetic case). For example, for a magnetic field scaling as $B \propto \rho^{1/2}$, the compression factor $\rho_2/\rho_1 \approx 0.16 \left[\beta/\left(\beta + 1 \right) \right]\,\mathscr{M}^2$ \citep{molina12,federrath10}, where $\beta$ is the ratio of gas pressure over magnetic pressure in the pre-shock region. In the case of the X-1 bubble, we will see later (Section 3.3.3) that for a typical sound speed $\sim$10 km s$^{-1}$ in the ambient medium, the Mach number of the shock is $\sim$10--15. Therefore, we expect a post-shock compression factor of $\sim$ (1 to a few) $\times$ 10 (that is, $n_e \lesssim$ a few $\times 10$ cm$^{-3}$). This is indeed consistent with the density measured from the [S {\footnotesize{II}}]$\lambda 6716/6731$ doublet.

Thirdly, we measured the index $R \equiv \left({\mathrm{[O {\footnotesize{III}}]}}\lambda 4959 + {\mathrm{[O {\footnotesize{III}}]}}\lambda 5007\right)/{\mathrm{[O {\footnotesize{III}}]}}\lambda 4363$. This is a standard indicator of the electron temperature $T_e$ \citep{osterbrock06}. Nebulae ionized by stellar continua and/or by X-ray emission have typical $T_e \approx 10,000$ K, while a temperature $T_e \gtrsim 20,000$ K typically identifies collisionally ionized gas. In our case, we find: $R \approx 16 \pm 2$ for the ULX bubble in Obs1, which corresponds to $T_e \approx (45,000 \pm 3000)$ K; $R \approx 18 \pm 2$ for the ULX bubble in Obs2, which corresponds to $T_e \approx (42,000 \pm 3000)$ K; and $R \approx 240 \pm 60$ for the H {\footnotesize{II}} region, which corresponds to $T_e \approx (9700 \pm 600)$ K. This is consistent with our interpretation of the two regions.

The intensity ratio between [O {\footnotesize{III}}]$\lambda 5007$ and H$\beta$ shows (Figure 10, bottom panel) peaks at the edge of the ULX bubble and in the H {\footnotesize{II}} region, but has its lowest value inside the bubble, close to the ULX position. X-ray photo-ionization by a central source, on the other hand, would result in higher excitation, {\it i.e.}, higher 
[O {\footnotesize{III}}]$\lambda 5007$/H$\beta$ ratios, close to the ionizing source. This is indeed observed in several other ULX bubbles, suggesting that both shock- and photo-ionisation play a role in those objects. However, for the NGC\,5585 X-1 bubble, shock-excitation appears to be by far the dominant mechanism.

Another useful intensity ratio often used in the literature is [O {\footnotesize{II}}]$\lambda\lambda 3727,3729$ over [O {\footnotesize{III}}]$\lambda 5007$. This ratio often suffers more from  uncertainties in the intrinsic extinction; however, it provides a clean distinction between X-ray photoionization and collisional ionization. Specifically, values of $I(3727,3729)/I(5007) \sim 1$--10 indicate shocks, while $I(3727,3729)/I(5007) < 1$ are exclusive of a power-law ionizing flux \citep{baldwin81,moy02}. In this ULX bubble, we measure $I(3727,3729)/I(5007) \approx 6.7$ ($\approx$7 after de-reddening), near the upper range of the shock-ionization regime.

A list of the observed intensities of the most important lines significantly detected in the LBT spectra is reported in Table 5, expressed as a ratio to the H$\beta$ intensity ($= 100$); they are divided into the three spatially distinct components described earlier (collisionally ionized ULX bubble along the minor axis and along the major axis, and photo-ionized H {\footnotesize{II}} region). We also list in Table 5 the average intensity of the H$\alpha$ and H$\beta$ emission measured along the slit, in units of erg cm$^{-2}$ s$^{-1}$ arcsec$^{-2}$. The total fluxes and luminosities of those two lines (and of the other lines, via their line ratios) can be estimated from those values, multiplied by the projected sky area of the bubble and of the H {\footnotesize{II}} region. We verified that those estimates are consistent with the direct measurements of narrow-band fluxes from the {\it HST} images (Section 3.3.6), and with the values measured for this bubble by \cite{matonick97}.

\begin{figure}
\centering
\vspace{-0.8cm}
\includegraphics[height=0.405\textwidth, angle=0]{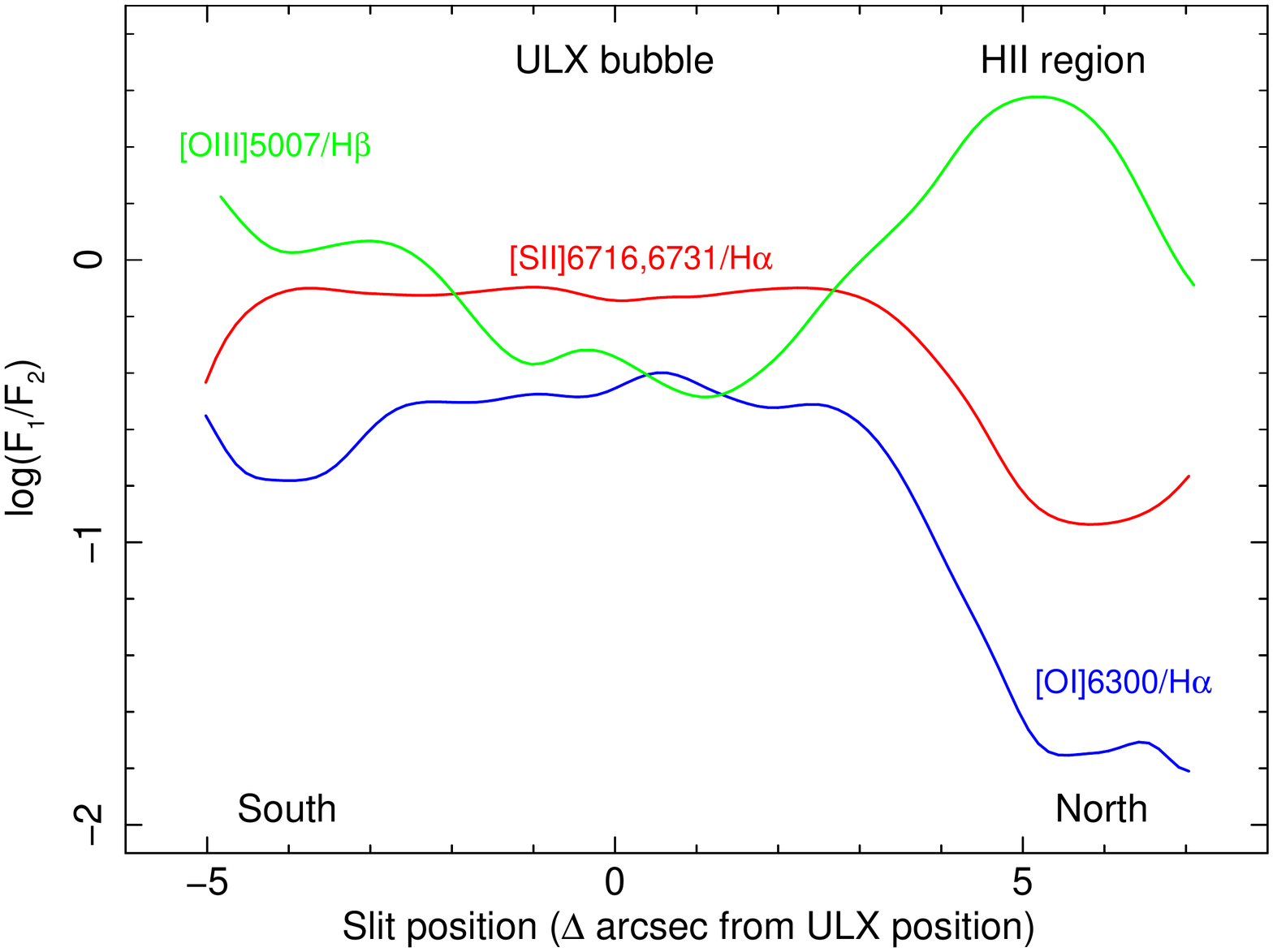}\\[-20pt]
\includegraphics[height=0.405\textwidth, angle=0]{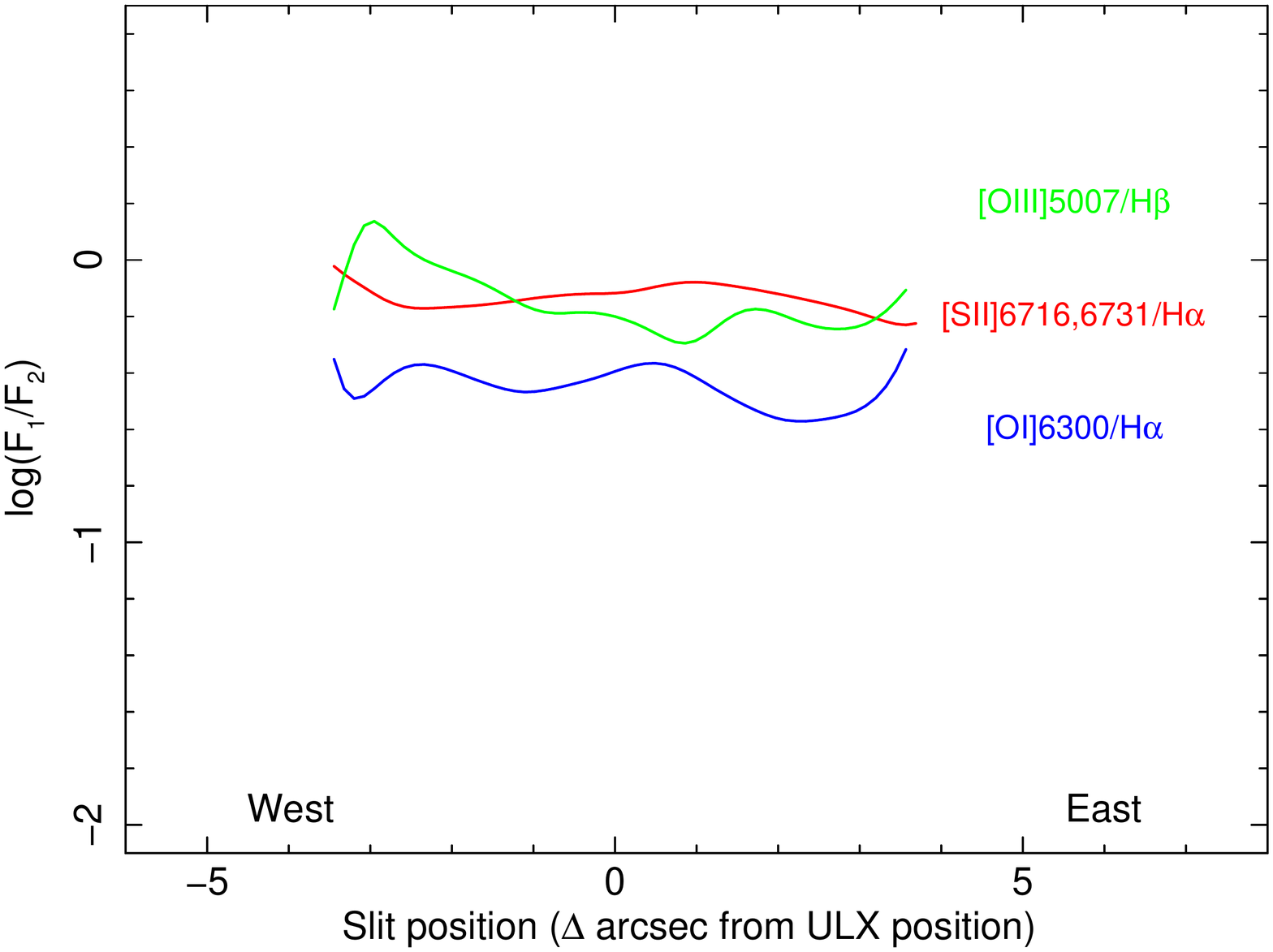}
\vspace{-0.4cm}
 \caption{
 Top panel: diagnostic line ratios [S II]/H$\alpha$ (red), [O I]/H$\alpha$ (blue) and [O III]/H$\beta$ (green), plotted along the slit in the Obs1 position ({\it i.e.}, along the major axis of the bubble, from south to north; see also Figure 4). All curves were smoothed using a Gaussian kernel with a 3-pixel sigma  ($\approx$0$^{\prime\prime}$.4). The drop in the flux ratios [S II]/H$\alpha$ and [O I]/H$\alpha$, and the corresponding increase in [O III]/H$\beta$, mark the transition between shock-ionized ULX bubble and photo-ionized H II region, $\approx$4$^{\prime\prime}$ north of X-1 (but partially overlapping).
 Bottom panel: as in the top panel, but along the slit in the Obs2 position, from west to east.
 }
  \label{lbtred}
\end{figure}

\begin{figure*}
\centering
\includegraphics[width=0.49\textwidth, angle=0]{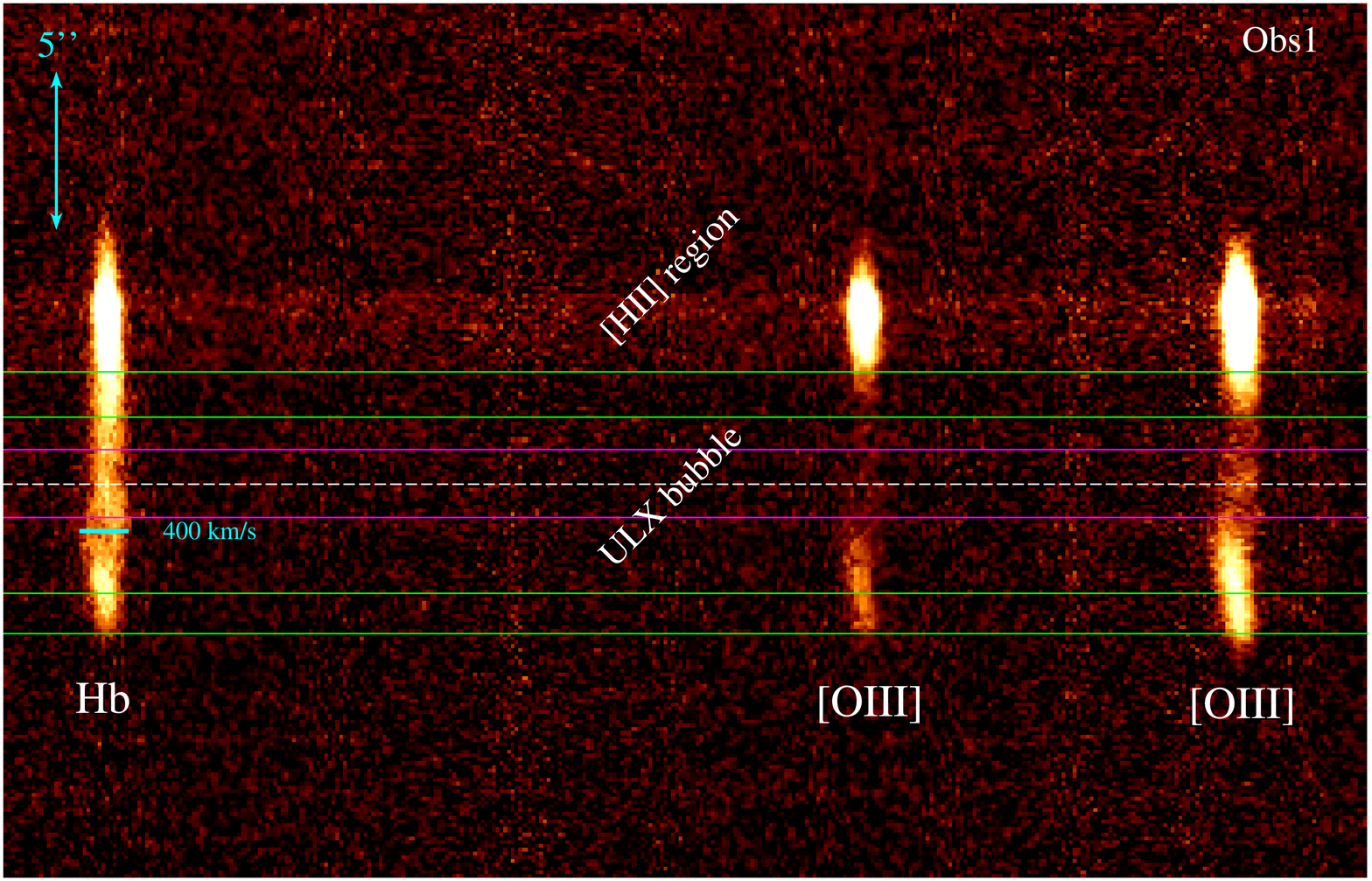}
\includegraphics[width=0.49\textwidth, angle=0]{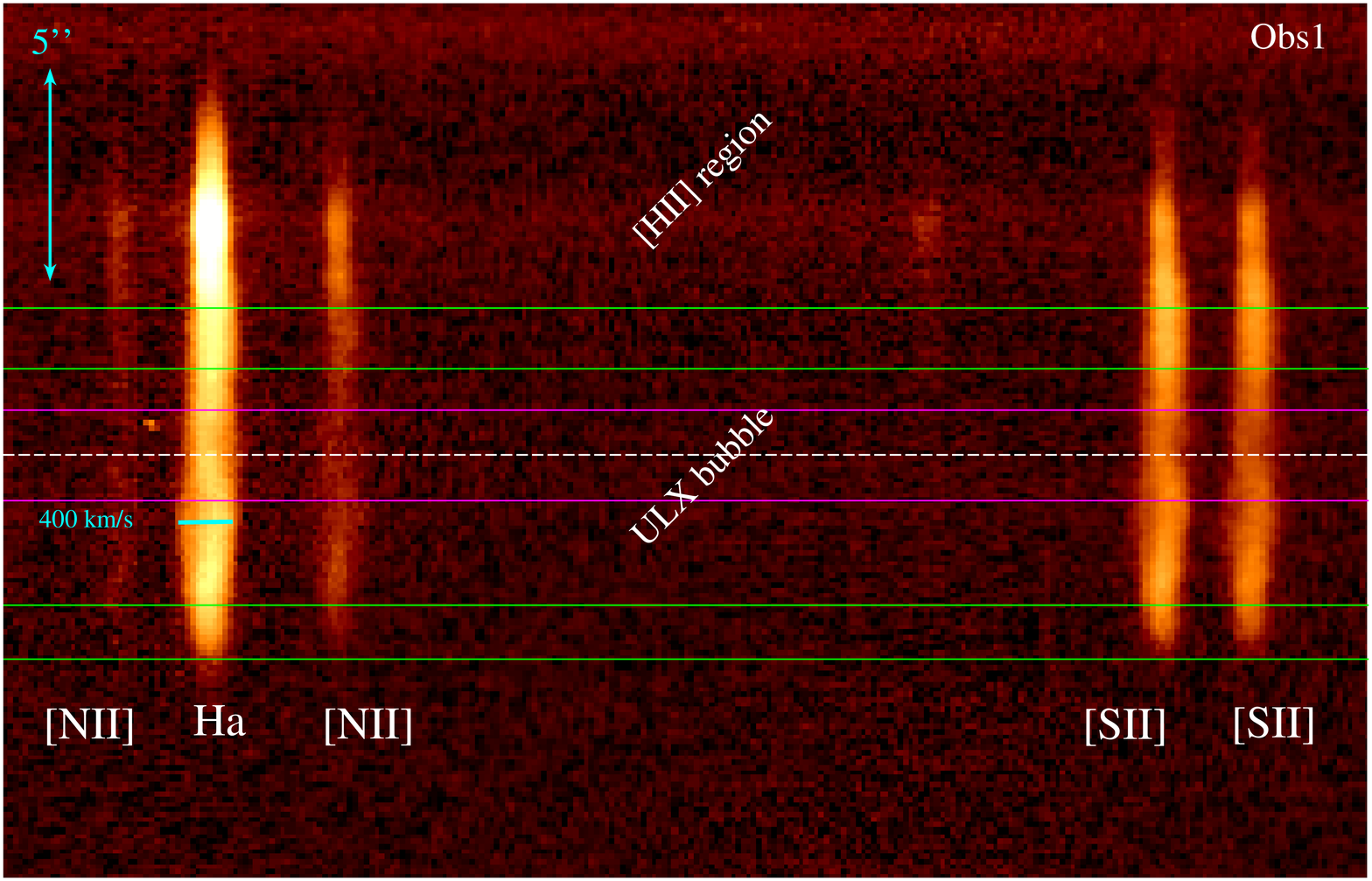}\\
\includegraphics[width=0.49\textwidth, angle=0]{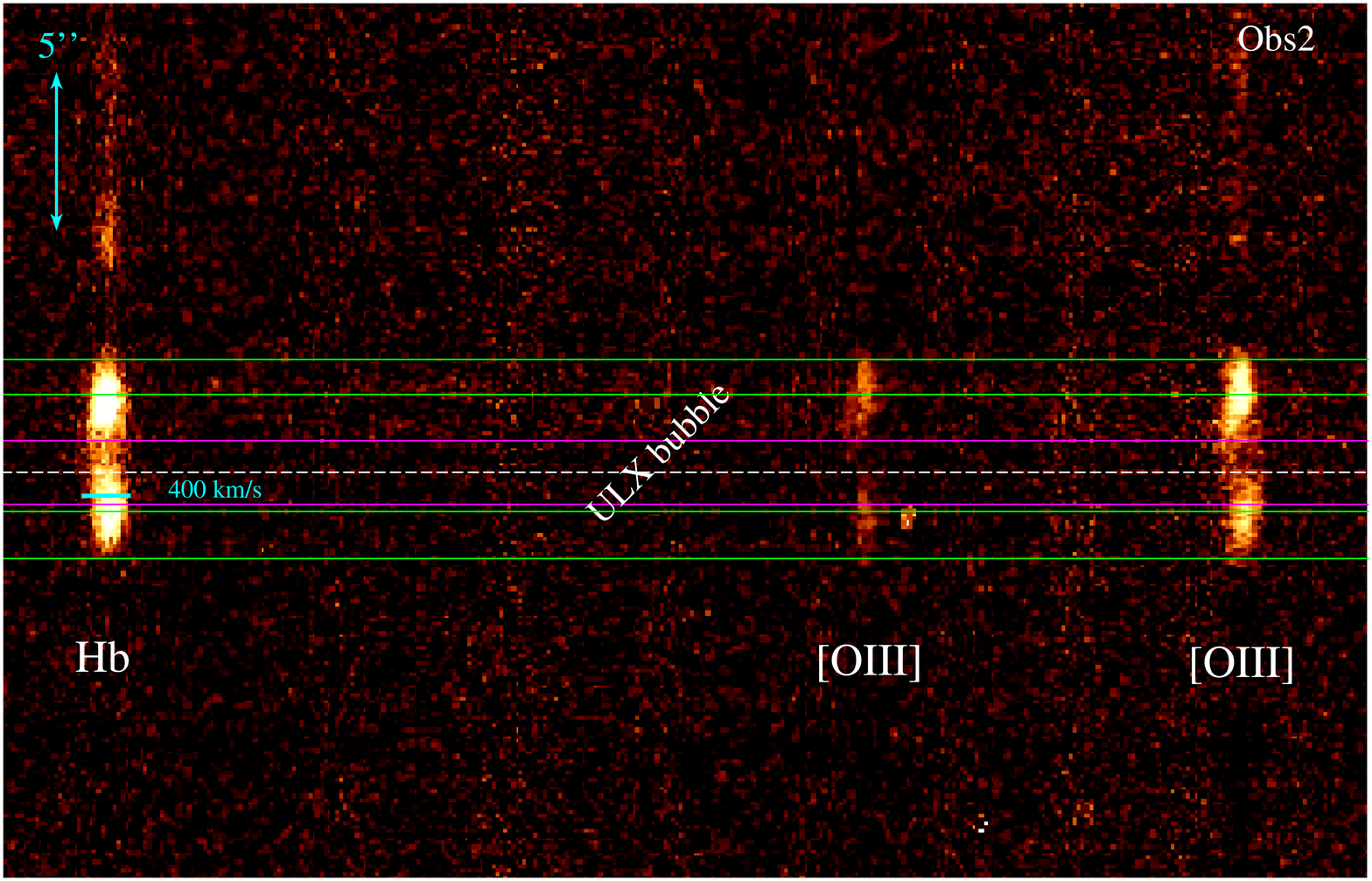}
\includegraphics[width=0.49\textwidth, angle=0]{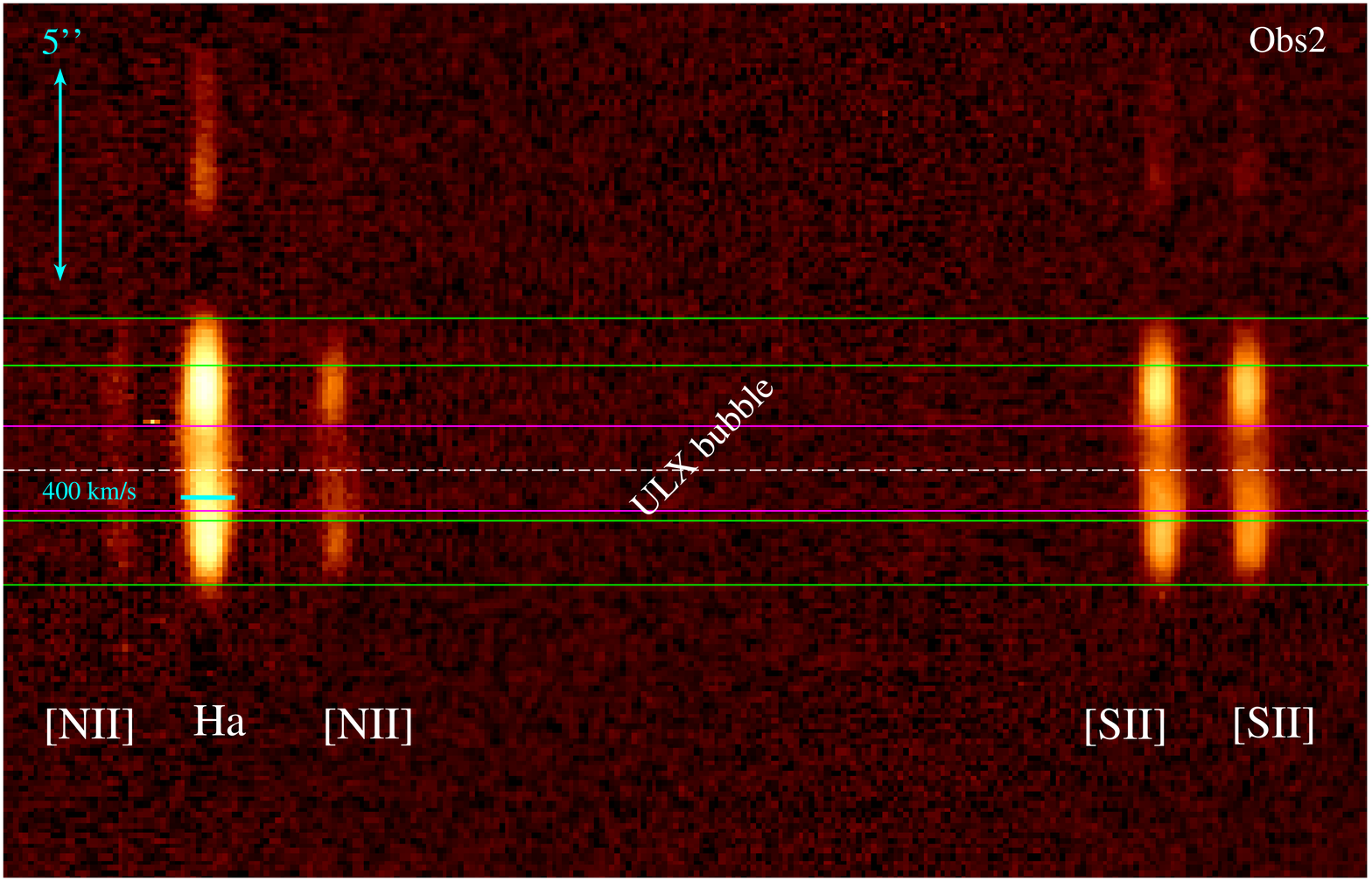}
 \caption{
Sky-subtracted, 2-dimensional LBT spectra in the Obs1 (top panels) and Obs2 (bottom panels) slit positions. In the top panels, north is up; in the bottom panels, east is up. The left panels illustrate the region between H$\beta$ and [O III] $\lambda 5007$; the right panels are for the region between H$\alpha$ and [S II]$\lambda 6731$ (notice the different zoom factor of left and right panels, for graphical purposes only). Parallel magenta lines define the ``central'' location of the bubble, defined as $\pm1^{\prime\prime}$ either side of the location of X-1 along both sits; this is the central region used in Section 3.3.3 to estimate $v_{\rm s}$. The dashed white line is the location of X-1. A 400 km s$^{-1}$ horizontal bar is overplotted to highlight the observed line broadening.
Parallel solid green lines mark four spectral regions at the four edges of the bubble (defined in Section 3.3.3), which we chose for further spatially resolved analysis of FWHMs and Doppler shifts of the lines (Sections 3.3.3, 3.3.4). Notice also that in the Obs1 position, the H II region is well visible (and partly overlapping) at the top end of the ULX bubble, while of course it is not on the slit in the Obs2 position.
 }
  \label{twodspectra}
\end{figure*}

\subsubsection{Shock velocity from optical line width}
Intrinsic broadening of the emission lines is another strong clue that a nebula is shock-ionized rather than photo-ionized; this is for example a classic selection criterion between SNRs and H {\footnotesize{II}} regions \citep{points19}. In our LBT study, we used the intrinsic full-width-half-maximum (FWHM$_{\rm int}$) of the emission lines from collisionally ionized plasma as a proxy for the shock velocity $v_{\rm s}$. Even for the simple case of a spherical bubble expanding in a uniform medium, the relation between these two quantities depends on whether the shocks are adiabatic or fully radiative, and on what portion of the bubble is observed. We discuss and quantify this relation in Appendix A; the take-away message of this analysis is that $v_{\rm s}$ is expected to be $\sim$0.5 times the FWHM measured from the central region of the bubble, and $\sim$3/4 of the average FWHM integrated over the whole bubble.  

To obtain the FWHM$_{\rm int}$ of the main lines from the bubble region, we need to subtract the instrumental width FWHM$_{\rm ins}$ from the observed width FWHM$_{\rm obs}$, using the relation FWHM$_{\rm int} = ({\mathrm{FWHM}}^2_{\rm obs} - {\mathrm{FWHM}}^2_{\rm ins})^{1/2}$.
We used the narrow emission lines from the photo-ionized [H {\footnotesize{II}}] region north of the ULX bubble to determine the instrumental line width in our blue and red LBT spectra, separately for each night. For the blue spectra, 
we found an instrumental width 
FWHM$_{\rm ins,b} \approx 2.05$ \AA\ on the first night and FWHM$_{\rm ins,b} \approx 2.13$ \AA\ on the second night; for the red spectra, we found FWHM$_{\rm ins,r} \approx 3.40$ \AA\ and FWHM$_{\rm ins,r} \approx 3.41$ \AA\ on the two nights. 
We then measured the FWHM$_{\rm obs}$ of H$\gamma$, H$\beta$ and 
[O {\footnotesize{III}}]$\lambda 5007$ in the blue spectra, and of H$\alpha$, [S {\footnotesize{II}}]$\lambda$6716 and [S {\footnotesize{II}}]$\lambda$6731 in the red spectra, in order to determine their FWHM$_{\rm int}$. We converted the width of each line into velocity units and computed an average values between those six lines.

More specifically, we measured the FWHM$_{\rm obs}$ of those strong lines separately in two ways: i) from the central section of the ULX bubble in both slits (the section of the 2-dimensional spectra between the parallel magenta lines overplotted in Figure 11, defined as $\pm 1^{\prime\prime}$ around the ULX position); ii) integrated over the whole ULX bubble region in both slits. For the central region, after correcting for the instrumental broadening, and averaging over the two slits and the two nights of observations, we find an average FWHM$_{\rm int,c} \approx 210 \pm 20$ km s$^{-1}$. From the relations discussed in Appendix A, this suggests $v_{\rm s} \sim 100 \pm 10$ km s$^{-1}$. For the full slit spectra (which is a proxy for the spectrum of the full bubble), we find instead an average FWHM$_{\rm int,T} \approx 170 \pm 30$ km s$^{-1}$, which corresponds to $v_{\rm s} \sim 125 \pm 20$ km s$^{-1}$ for a fully radiative shock. We conclude that the most likely value of $v_{\rm s}$ is between $\sim$100--125 km s$^{-1}$. Henceforth, for the purpose of energy budget and characteristic mechanical power of the bubble, we will take a characteristic $v_{\rm s} \approx 125$ km s$^{-1}$, but our main conclusions would remain essentially valid also for $v_{\rm s} \approx 100$ km s$^{-1}$.

Finally, we extracted spatially resolved spectra also at the outer edges of the bubble region in the two slits (between the parallel green lines overplotted in Figure 11), to measure the line broadening near the northern, southern, eastern and western edges. We defined the four edge regions as follow: 
along the Obs 1 slit, between $\approx -4^{\prime\prime}.8$ and $-3^{\prime\prime}.5$ to the south, and between $\approx$ $+2^{\prime\prime}.0$ and $+3^{\prime\prime}.4$ to the north of X-1; along the Obs 2 slit, between $\approx -2^{\prime\prime}.7$ and $-1^{\prime\prime}.2$ to the west, and between $\approx$ $+2^{\prime\prime}.5$ and $+3^{\prime\prime}.5$ to the east of X-1.
The intensity, velocity and FWHM profiles along the slit of the bright H$\alpha$ line were used to define and select those edge intervals. Over those regions, the velocity and FWHM of the lines are relatively stable, and therefore we consider them representative of the actual conditions at the expanding front.

We find that the FWHM$_{\rm int}$ varies between $\approx$110--150 km s$^{-1}$ (averaged over the strongest lines) in the four regions (Figure 12). In principle, the projected velocity exactly at the edge of a radially expanding bubble should be null; however, we are extracting the spectra over finite spatial regions. Moreover, we also see in other bubbles (NGC\,7793-S26: M.W. Pakull et al., in prep.) that there is substantial line broadening (of order of the shock velocity) where the outflow impacts the outer shell. This may be evidence of sideways splash or backflow of some of the shocked gas.

\subsubsection{Systemic velocity and Doppler shifts}
We fitted the strongest lines in the red spectra\footnote{We only used H$\alpha$ and the [S II] doublet for the redshift analysis, because the wavelength calibration of the blue spectra has a larger uncertainty, due to the lack of suitable arc lines in that region.} with Gaussians, to measure their central positions, and determined an average systemic velocity of $190 \pm 5$ km s$^{-1}$ over the whole bubble (average of both slits and both nights), in the heliocentric reference frame. For the [H {\footnotesize{II}}] region, we measured a heliocentric systemic velocity $\approx$200 $\pm 10$ km s$^{-1}$. The heliocentric velocity of NGC\,5585 is $293 \pm 3$ km s$^{-1}$ \citep{epinat08}, but X-1 is located on the approaching side of the rotating galactic disk, where the projected rotational velocity is $\approx$ $-70$ km s$^{-1}$ \citep{cote91,epinat08}. The fact that X-1 is still located approximately at the centre of its own bubble, after an expansion age of $\approx$6 $\times 10^5$ yr (Section 4.1), suggests that the compact object was born with a low kick velocity: we estimate a projected velocity in the plane of the sky $\lesssim$50 km s$^{-1}$.  

Furthermore, there are significant velocity differences between different regions of the ULX bubble. From our separate spectral analysis at the four edges of the bubble, we find that the ionized gas at the northern edge is receding at a projected speed $\approx$30 km s$^{-1}$ higher than at the southern edge (Figure 12); conversely, the gas at the eastern edge is moving away $\approx$25 km s$^{-1}$ more slowly than the gas at the western edge and $\approx$30 km s$^{-1}$ more slowly than the gas at the southern edge. In the spectra from the central region, the average recession velocity is $195 \pm 5$ km s$^{-1}$.



\subsubsection{Reddening}
We examined the Balmer decrements to obtain a more accurate estimate of the total reddening (Milky Way plus NGC\,5585 halo) in front of the bubble and of the H {\footnotesize{II}} region. 
First, we considered the bubble region. The theoretical intensity ratios (without reddening) for collisionally-ionized gas depend only weakly on the shock velocity and metal abundance. Assuming $100 \lesssim v_{\rm s} \lesssim 150$ km s$^{-1}$ (Section 3.3.3) and sub-solar (Large Magellanic Cloud) metal abundance (justified by our subsequent analysis in Section 3.3.7 and by the general properties of NGC\,5585), from the {\sc mappings iii} tables of \cite{allen08}, we expect: 
$I({\rm{H}}\alpha)/I({\rm{H}}\beta) \approx 3.04$--3.22 (with a central value of $I({\rm{H}}\alpha)/I({\rm{H}}\beta) \approx 3.14$ for $v_{\rm s} = 125$ km s$^{-1}$);  
$I({\rm{H}}\gamma)/I({\rm{H}}\beta) \approx 0.45$--0.46;
$I({\rm{H}}\delta)/I({\rm{H}}\beta) \approx 0.25$.
The observed Balmer flux ratios are:   
$F({\rm{H}}\alpha)/F({\rm{H}}\beta) = 3.42 \pm 0.24$;  
$F({\rm{H}}\gamma)/F({\rm{H}}\beta) = 0.46 \pm 0.03$;
$F({\rm{H}}\delta)/F({\rm{H}}\beta) = 0.27 \pm 0.03$ 
(Table 5, from the average of the Obs1 and Obs2 bubble-region values). For the differential reddening between the lines, we used the extinction curve parameterized by \cite{esteban14}. From $F({\rm{H}}\alpha)/F({\rm{H}}\beta)$ we obtain  $E(B-V)_{\alpha/\beta} = 0.09 \pm 0.08$ mag, $E(B-V)_{\gamma/\beta} \lesssim 0.07$ mag, 
$E(B-V)_{\delta/\beta} < 0.08$ mag.
Taking an average of the three measurements, 
we estimate an intrinsic $E(B-V) = 0.02^{+0.04}_{-0.02}$ mag, in addition to the Galactic foreground component $E(B-V) \approx 0.014$ mag. Given the small value, low significance and large relative uncertainty of the intrinsic component, in the rest of this paper we will generally correct only for the foreground Galactic extinction, unless explicitly mentioned.

We then used the same Balmer-decrement method to estimate the reddening
in the H {\footnotesize{II}} region. Here we assumed Case-B recombination, $T_e = 10,000$ K, low-density limit \citep{osterbrock06}; the theoretical flux ratios for photo-ionized gas are: 
$I({\rm{H}}\alpha)/I({\rm{H}}\beta) = 2.87$;  
$I({\rm{H}}\gamma)/I({\rm{H}}\beta) = 0.468$;
$I({\rm{H}}\delta)/I({\rm{H}}\beta) = 0.259$. 
In our Obs1 spectrum, we chose H$\alpha$ and H$\delta$ for an estimate of the Balmer decrement (the measured value of H$\gamma$ is an outlier). Comparing the observed (reddened) values of $F({\rm{H}}\alpha)/F({\rm{H}}\beta) \approx 3.9$ and $F({\rm{H}}\delta)/F({\rm{H}}\beta) \approx 0.21$ (Table 5) with the theoretical ratios, we obtain $E(B-V)_{\alpha/\beta} \approx 0.30$ mag and  $E(B-V)_{\delta/\beta} \approx 0.36$ mag. Taking an average of those two values, we obtain a best estimate of the total reddening  $E(B-V) \approx 0.33$ mag.



\begin{figure}
\centering
\includegraphics[width=0.47\textwidth, angle=0]{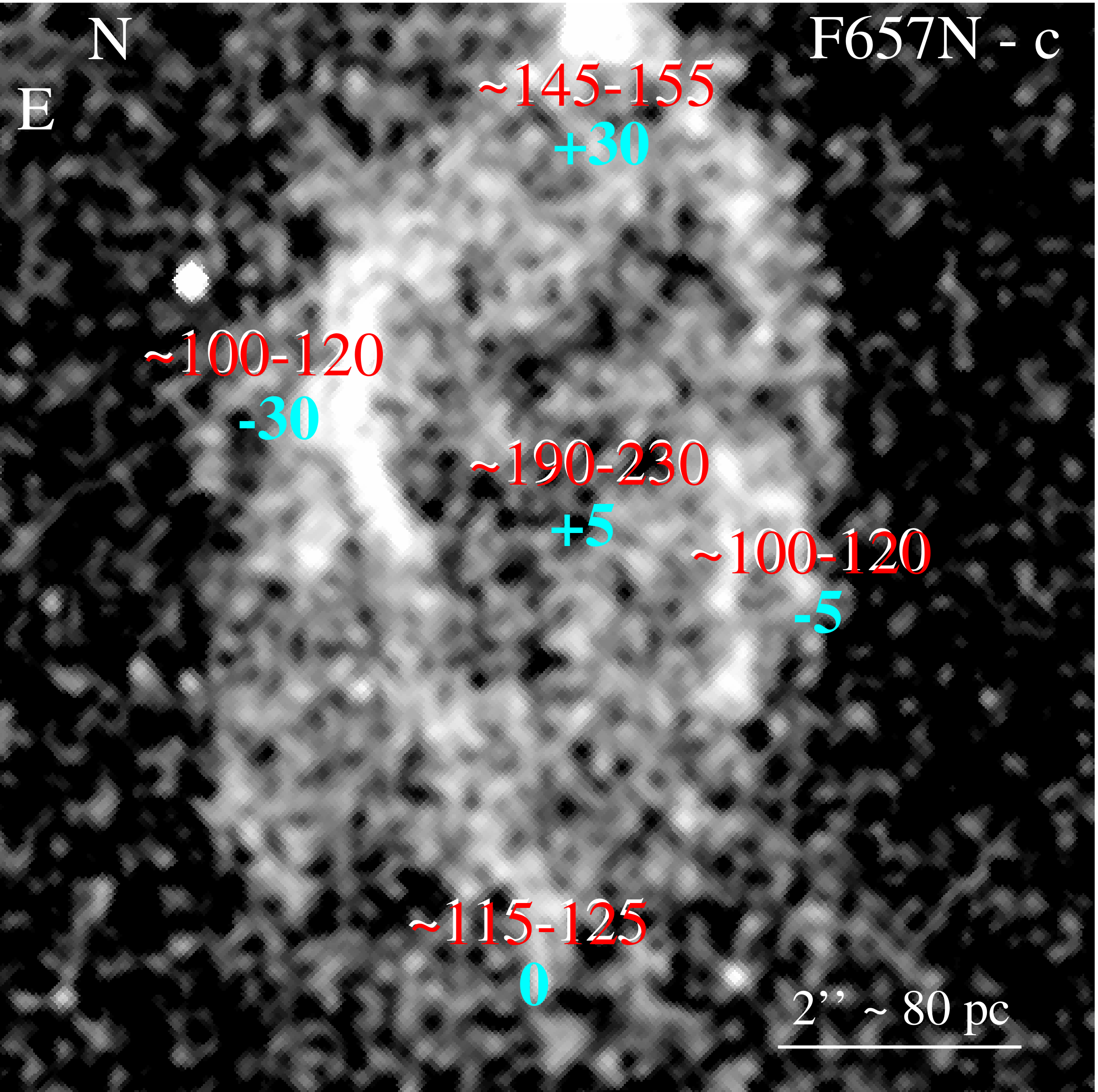}
 \caption{
 Approximate velocity parameters from an average of the strongest emission lines, superposed on a continuum-subtracted, Gaussian-smoothed image of the ULX bubble in the F657N filter. Red numbers represent the approximate range of intrinsic FWHMs (in km s$^{-1}$) measured in five characteristic sections of the bubble, from Gaussian fits to the LBT profiles of H$\alpha$, H$\beta$, H$\gamma$, [O III] $\lambda 5007$, [S II] $\lambda 6716$ and [S II]$\lambda 6731$. Blue numbers represent the approximate Doppler shift of the emission lines in the same four regions, with respect to the average systemic velocity over the ULX bubble region. The Doppler shifts are measured from the average shift of the central positions of H$\alpha$ and [S II] $\lambda 6716$ only; the reference average systemic velocity is $190 \pm 5$ km s$^{-1}$. 
 }
  \label{velocities}
\end{figure}

\subsubsection{Fluxes and luminosities}

The {\it HST} image in the F657N band (Table 6) provides an accurate total flux and luminosity for the combined emission of H$\alpha$, [N {\footnotesize{II}}]$\lambda6584$ and [N {\footnotesize{II}}]$\lambda6548$ in the ULX bubble region. We combined that with the line ratios inferred from the LBT spectra (Table 5), and determined the emitted luminosity of H$\alpha$ and other lines. We obtain $L_{\rm{H}\alpha} = (9.7 \pm 0.7) \times 10^{37}$ erg s$^{-1}$ and hence (from the theoretical line ratios) $L_{\rm{H}\beta} = (3.2 \pm 0.2) \times 10^{37}$ erg s$^{-1}$, where we have corrected only for the foreground Galactic reddening $E(B-V) = 0.014$ mag.  If we apply also a (speculative) intrinsic reddening $E(B-V) \approx 0.02$ mag, the emitted Balmer line luminosity increases to $L_{\rm{H}\alpha} \approx 10.2 \times 10^{37}$ erg s$^{-1}$ and $L_{\rm{H}\beta} \approx 3.4 \times 10^{37}$ erg s$^{-1}$. Another important line luminosity that we will use as a proxy for the kinetic energy is the one from [Fe {\footnotesize{II}}]$\lambda1.64 \mu$m (F164N filter), for which we infer $L_{\rm{[FeII]}} = (1.1 \pm 0.1) \times 10^{37}$ erg s$^{-1}$ (Table 6); the additional intrinsic reddening would only increase this value by $\approx$1\%).


A detailed analysis of the line and continuum emission from the H {\footnotesize {II}} region is beyond the scope of this work. Here we report only a few results that may help characterize the environment around the ULX.  We determined a total luminosity (corrected for foreground Galactic extinction) of $\approx$3.6 $\times 10^{37}$ erg s$^{-1}$ in the F657N filter, and we estimate that H$\alpha$ contributes $\approx$95\% of this emission ({\it i.e.}, $\approx$3.4 $\times 10^{37}$ erg s$^{-1}$). Correcting for the additional intrinsic extinction (Section 3.3.4) $A_{\rm{H}\alpha} \approx 0.818 \times 3.1 \times 0.31$ mag, we infer an emitted luminosity $L_{\rm{H}\alpha} \approx 7 \times 10^{37}$ erg s$^{-1}$. The intrinsic reddening derived from the Balmer decrement is certainly an upper limit for the reddening of the continuum emission; the latter is dominated by bright stars that may have already cleared the gas around them. The continuum luminosity in the broad-band {\it HST} filters is summarized in Table 7.

We ran simulations with the {\sc starburst99} web-based software \citep{leitherer99,leitherer14} to estimate the stellar population properties consistent with such luminosities and colours. We find that both the H$\alpha$ luminosity and the optical colours are consistent with an instantaneous burst of star formation with a mass $M \approx (2000 \pm 500) M_{\odot}$, after an age $\lesssim 3.5$ Myr (the timescale after which O stars die, which makes the U-band and H$\alpha$ luminosities drop quickly). Alternatively, we find acceptable solutions for continuous star formation at a rate of $\approx$4--7 $\times 10^{-4} M_{\odot}$ yr$^{-1}$ at an age $\lesssim 4$ Myr.


\subsubsection{Metal abundance}
We determined the metal abundance independently in the H {\footnotesize{II}} region and in the ULX bubble. For the H {\footnotesize{II}} region, first we re-visited our temperature estimate (Section 3.3.3), now taking into account also the reddening derived in Section 3.3.4. We obtain an electron temperature in the [O {\footnotesize{III}}] zone $T_e({\mathrm{O {\footnotesize{III}}}}) = 10,600 \pm 400$ K. The corresponding temperature in the [O {\footnotesize{II}}] zone, $T_e({\mathrm{O {\footnotesize{II}}}})$ can be empirically obtained from the scaling relation of \cite{garnett92}, which gives $T_e({\mathrm{O {\footnotesize{II}}}}) = 10,400 \pm 400$ K.
With these two temperatures, and with the reddening-corrected intensity of the strong oxygen lines, we can then determine the oxygen abundance, using the relations of \cite{izotov06}. We obtain a metallicity index $12 + \log({\rm O}/{\rm H}) \approx 8.19$ dex. Using instead the calibration of \cite{pagel92} (see also \citealt{pilyugin05}), we obtain $12 + \log({\rm O}/{\rm H}) \approx 8.17$ dex. Thus, the metal abundance appears to be intermediate between those of the Large Magellanic Cloud ($12 + \log({\rm O}/{\rm H}) = 8.35 \pm 0.06$) and the Small Magellanic Cloud ($12 + \log({\rm O}/{\rm H}) = 8.03 \pm 0.10$) \citep{russell92}. An alternative calibration for the oxygen abundance in an H {\footnotesize{II}} region is based on the intensity ratio [O {\footnotesize{II}}]$\lambda 6583$/H$\alpha$ \citep{perez-montero09}. In our case, we measured [O {\footnotesize{II}}]$\lambda 6583$/H$\alpha$ $\approx 0.05$, which implies an abundance $12 + \log({\rm O}/{\rm H}) \approx 8.04$, consistent with the Small Magellanic Cloud.

For the ULX bubble, we used again the ratio [N {\footnotesize{II}}]$\lambda 6583$/H$\alpha$, which is a strong function of metallicity in shock-ionized gas \citep{allen08}. The empirical ratio in the bubble region is $\approx 0.14 \pm 0.02$ (Table 5), which is similar to the ratios ($\approx$0.10--0.11) predicted by the {\sc mappings iii} code for shock velocities in the range $\approx$100--150 km s$^{-1}$ at Large Magellanic Cloud metallicity. As a comparison, the same line ratio is predicted to be $\approx$0.04--0.05 at Small Magellanic Cloud metallicity, and $\approx$0.48--0.51 at solar metallicity. Thus, it is possible that the more evolved bubble region is also slightly more enriched than the currently star-forming region north of it, but they are both sub-solar.






\begin{table}
\caption{Observed relative intensity$^{a}$ of the main lines in the ULX bubble and H II regions, from the LBT spectra, and absolute average intensity of H$\beta$. Errors are $\sim$10\% for the weaker lines (those with an intensity $\lesssim$0.5 times H$\beta$), and $\sim$5\% for the stronger lines (those with an intensity $\gtrsim$ H$\beta$).}
\vspace{-0.4cm}
\begin{center}
\begin{tabular}{lccc}  
\hline \hline\\[-5pt]    
Line &     \multicolumn{3}{c}{Extraction Region} \\[5pt]   
    &   ULX bubble$^{b}$   & ULX bubble$^c$   & 
    H {\footnotesize{II}} region$^{d}$\\ 
    &   Obs1 slit   & Obs2 slit   & 
    Obs1 slit\\
\hline  \\[-5pt]
& \multicolumn{3}{c}{{\it {Relative Intensities}}}\\
\hline  \\[-5pt]
[O {\footnotesize{II}}]$\lambda\lambda$3727,3729  &  502  &  469 & 115 \\ [3pt]
[Ne {\footnotesize{III}}]$\lambda$3869  &  20  &  26 & 16 \\ [3pt]
H${\zeta}\,\lambda$3889  &    &   & 10 \\ [3pt]
$^e$H${\epsilon}\,\lambda$3970  &  27  & 24  & 14\\ [3pt]
[S {\footnotesize{II}}]$\lambda$4069  &  18  &  11 &  \\ [3pt]
H${\delta}\,\lambda$4102  &  27  & 27  & 21\\ [3pt]
H${\gamma}\,\lambda$4340  &  48  & 43  & 34\\ [3pt]
[O {\footnotesize{III}}]$\lambda$4363  &  6  &  5 & 2 \\ [3pt]
H${\beta}\,\lambda$4861  &  100  & 100  & 100\\ [3pt]
[O {\footnotesize{III}}]$\lambda$4959  &  20  &  21 & 117 \\ [3pt]
[O {\footnotesize{III}}]$\lambda$5007  &  77  &  69 & 353 \\ [3pt]
He {\footnotesize{I}}$\lambda$5876  & 16   &  10 & 15 \\ [3pt]
[O {\footnotesize{I}}]$\lambda$6300  &  93  &  76 & 7 \\ [3pt]
[S {\footnotesize{III}}]$\lambda$6312  &    &   & 4 \\ [3pt]
[O {\footnotesize{I}}]$\lambda$6364  &  33  &  23 & 3 \\ [3pt]
[N {\footnotesize{II}}]$\lambda$6548  &  21  &  16 & 3 \\ [3pt]
H${\alpha}\,\lambda$6563  &  356  & 328  & 391\\ [3pt]
[N {\footnotesize{II}}]$\lambda$6583  &  55  &  44 & 19 \\ [3pt]
[S {\footnotesize{II}}]$\lambda$6716  &  153  &  138 & 33 \\ [3pt]
[S {\footnotesize{II}}]$\lambda$6731  &  104  &  96 & 22 \\ [3pt]
\hline\\[-5pt]
& \multicolumn{3}{c}{{\it {Average Intensity}$^f$}}\\
\hline  \\[-5pt]
H${\beta}\,\lambda$4861  &  $1.0\pm 0.1$  & $1.8\pm 0.1$  & $5.5\pm 0.1$\\ [3pt]
\hline
\vspace{-0.5cm}
\end{tabular} 
\end{center}
\begin{flushleft} 
\footnotesize{
$^a$: normalized to H$\beta$, without reddening corrections;\\
$^b$: ULX bubble spectrum extracted between $-4^{\prime\prime}.3$ and $+2^{\prime\prime}.0$ from the ULX position (positive values meaning to the north of the ULX), along the slit in the Obs1 position. The ULX bubble extends from $\approx -5^{\prime\prime}$ to $\approx +4^{\prime\prime}$ from the ULX position; however, we only used the slit section between $-4^{\prime\prime}.3$ and $+2^{\prime\prime}.0$ for the line intensities in this Table, to maximize signal-to-noise and avoid contamination from the  H {\footnotesize{II}} region.\\
$^c$: ULX bubble spectrum extracted from Obs2, between $-2^{\prime\prime}.7$ and $+3^{\prime\prime}.5$ from the ULX position (positive values meaning to the east);\\
$^d$: H {\footnotesize{II}} region spectrum extracted between $+4^{\prime\prime}.5$ and $+6^{\prime\prime}.2$ from the ULX position, along the slit in the Obs1 position;\\
$^e$: blended with [Ne III]$\lambda$3968 and Ca II $\lambda$3969;\\
$^f$: average intensity on the slit, in units of $10^{-16}$ erg cm$^{-2}$ s$^{-2}$ arcsec$^{-2}$. 
}
\end{flushleft}
\end{table}

\subsubsection{Another diagnostic line: [O I]$\lambda 6300$}
The low-ionization line [O {\footnotesize{I}}]$\lambda 6300$ is another indicator of the spatial boundary between the collisionally ionized ULX bubble and the photo-ionized H {\footnotesize{II}} region (Figure 10, top panel). Surprisingly, the average observed intensity ratio $I(6300)/I({\rm{H}}\alpha) \approx 0.25$ (Table 5) in the ULX bubble is much higher than predicted by the {\sc mappings iii} tables, which suggest instead values in the range $\approx$0.02--0.05 for a plausible grid of shock velocities (100 km s$^{-1}$, 125 km s$^{-1}$, 150 km s$^{-1}$), metallicities (Small Magellanic Cloud, Large Magellanic Cloud, solar) and magnetic fields ($B = 0$, $B = 1 \mu$G, $B = 3.23 \mu$G; the last value represents the field at thermal/magnetic pressure equipartition, for a gas density $n = 1$ cm$^{-3}$). Intensity ratios $\gtrsim$0.2 apparently require shock velocities $\gtrsim$600 km s$^{-1}$ according to the {\sc mappings iii} code; this is clearly ruled out by the rest of our data. On the other hand, other widely used shock ionization models \citep{cox85,hartigan87,hartigan94} do predict intensity ratios in our observed range; for example, a ratio $\approx$0.3 for a shock velocity $\approx$140 km s$^{-1}$ into fully neutral material \citep{cox85}. As an observational comparison, values of $I(6300)/I({\rm{H}}\alpha) \approx 0.25$--0.30 were found across the collisionally ionized Galactic supernova remnant W44, for which the estimated shock velocity is $\approx$110--150 km s$^{-1}$ \citep{mavromatakis03}.

\subsection{ULX radio bubble}

Our VLA observations reveal for the first time a bright radio nebula (size of $350 \times 230$ pc) associated with the optical/infrared ULX bubble (Figure 2). We measured an integrated flux density $S_{\nu} = 1.4 \pm 0.1$ mJy at 1.5 GHz. At the assumed distance of 8 Mpc, this corresponds to a 1.5-GHz luminosity density $L_\nu = (1.1 \pm 0.1) \times 10^{26}$ erg s$^{-1}$ Hz$^{-1}$ and a luminosity $L \equiv \nu L_\nu = (1.6 \pm 0.1) \times 10^{35}$ erg s$^{-1}$. Assuming an optically thin synchrotron spectrum of index $\alpha = -0.7$ (with $S_\nu \propto \nu^{\alpha}$), this corresponds to a 5-GHz luminosity $L = (2.3 \pm 0.2) \times 10^{35}$ erg s$^{-1}$. As a comparison, the radio bubble associated with S26 in NGC\,7793 (prototypical example of a jet-powered bubble) has a size of $\approx$300 $\times$ 150 pc and a 5-GHz luminosity $L = (1.6 \pm 0.1) \times 10^{35}$ erg s$^{-1}$. 

Although synchrotron emission is likely the dominant component, the strong Balmer line emission observed from the bubble suggests that there might also be a contribution to the radio continuum from free-free emission. However, using the ratio of H$\beta$ and radio emissivities from \cite{caplan86}, at a temperature of $\sim$40,000 K, we verified that free-free emission is expected to contribute only $\approx$2\% of the observed radio luminosity.

The radio nebula shows three regions of significantly enhanced emission (Figure 2). Two of them are consistent with knots or hot spots along the major axis (that is, along the putative direction of a jet, by analogy with S26). The third one appears to coincide with the shock front on the eastern side of the bubble, which is also the brightest in the optical/infrared lines, perhaps because the shock is advancing into an ambient medium with higher density. 

\begin{table}
\caption{Continuum-subtracted narrow-band emission from the ULX bubble}
\vspace{-0.4cm}
\begin{center}
\begin{tabular}{lccc}  
\hline \hline\\[-5pt]    
Filter/Line &  Exp.~Time &  Flux  & Luminosity$^a$\\
 &  (s)  & (erg cm$^{-2}$ s$^{-1}$) & (erg s$^{-1}$)\\
\hline  \\[-5pt]
F502N$^b$ & 2012 &  $(3.3 \pm 0.6) \times 10^{-15}$ & $(2.7 \pm 0.5) \times 10^{37}$ \\ [3pt]
F657N$^c$ & 2916  &  $(14.4 \pm 1.0) \times 10^{-15}$ & $(11.4 \pm 0.8) \times 10^{37}$  \\ [3pt] 
F673N$^d$ & 2185  &  $(8.5 \pm 0.5) \times 10^{-15}$ & $(6.8 \pm 0.4) \times 10^{37}$  \\ [3pt]
F164N$^e$ & 1659  &  $(1.5 \pm 0.1) \times 10^{-15}$ & $(1.1 \pm 0.1) \times 10^{37}$  \\ [3pt]
\hline 
\vspace{-0.5cm}
\end{tabular} 
\end{center}
\begin{flushleft} 
\footnotesize{
$^a$: corrected for foreground Galactic reddening $E(B-V) = 0.014$ mag;\\
$^b$: Covering [O {\footnotesize{III}}]$\lambda$5007;\\
$^c$: Covering H$\alpha$, [N {\footnotesize{II}}]$\lambda$6548,6583. H$\alpha$ contributes $\approx$85\% of the flux and luminosity in this filter, based on the LBT line ratios;\\
$^d$: Covering [S {\footnotesize{II}}]$\lambda$6716,6731;\\
$^e$: Covering [Fe {\footnotesize{II}}]$\lambda$16,440.}
\end{flushleft}
\end{table}

\begin{table}
\caption{Continuum brightness of the star clusters in the H II region.}
\vspace{-0.4cm}
\begin{center}
\begin{tabular}{lccc}  
\hline \hline\\[-5pt]    
Filter &  Exp.~Time &   Apparent Brightness  & Absolute Brightness$^a$\\
 &  (s)  & (mag) & (mag)\\
\hline  \\[-5pt]
F336W & 1200 &  $20.1 \pm 0.04$ & $-9.5 \pm 0.1$ \\ [3pt]
F555W & 1245  &  $21.4 \pm 0.03$ & $-8.2 \pm 0.1$ \\ [3pt] 
F814W & 1350  &  $21.3 \pm 0.10$ & $-8.3 \pm 0.1$ \\ [3pt]
\hline
\vspace{-0.5cm}
\end{tabular} 
\end{center}
\begin{flushleft} 
\footnotesize{
$^a$: corrected for foreground Galactic reddening $E(B-V) = 0.014$ mag.}
\end{flushleft}
\end{table}

\section{Discussion}

\subsection{Jet power derived from the ULX bubble}
Following \cite{pakull10}, we estimate the kinetic power of the bubble in two ways. First, we use a well known relation from standard bubble theory \citep{weaver77}, between kinetic power $P_{\rm w}$, size $R$ and age $t$ of the bubble, and mass density $\rho_0$ of the undisturbed interstellar medium:
\begin{equation}
    R \approx 0.76 \left(P_{\rm w}/\rho_0\right)^{1/5} t^{3/5}.
\end{equation}
We define $R_2 \equiv R/(100~{\rm pc})$; $t_5 \equiv t/(10^5~{\rm {yr}})$; $P_{39} \equiv P_{\rm w}/(10^{39}~{\rm{erg~s}}^{-1})$. We also assume the bubble is expanding into mostly neutral ambient gas, for which the mean molecular weight $\mu \approx 1.30$, and $\rho_0 \equiv \mu m_{\rm p} n_0 \approx 2.17 \times 10^{-24} n_0$ g cm$^{-3}$, with $n_0$ being the atomic number density of the unshocked medium. Thus, the previous scaling relation can be re-written as
\begin{equation}
    R_2 \approx 0.265 \left(P_{39}/n_0\right)^{1/5} t_5^{3/5}
\end{equation}
with $n_0$ in units of cm$^{-3}$.
We adopt an average bubble radius $R \approx 130$ pc, and we estimate the age of the bubble as $t = (3/5)\,R/v_{\rm s}$ where we identify the expansion speed with the shock velocity (Section 3.3.3). Thus, 
\begin{equation}
    t_5 = 5.9 R_2/v_2 \approx 6.1
\end{equation}
with $v_2 \equiv v_{\rm s}/(100~{\rm{km~s}^{-1}})$ and $v_{\rm s} \approx 125$ km s$^{-1}$. Alternatively, we can take $R$ as the semi-minor axis ($R_2 \approx 1.1$) and $v_{\rm s} \approx 100$ km s$^{-1}$ as estimated for the central region of the bubble: this gives an age $t_5 \approx 6.5$. Then, Equation (1) transforms to:
\begin{equation}
    \left(P_{39}/n_0\right) \approx 7.7 \times 10^2 \, R_2^5 \, t_5^{-3} \approx 13,
\end{equation}
with an uncertainty of a factor of 2, for a plausible range of shock velocities. 

The hydrogen number density $n_0$ of the interstellar gas can be estimated from the shock velocity and the radiative flux of the H$\beta$ line\footnote{We consider only the main shock contribution to the line luminosity, and not the precursor, because we have already shown (Section 3.3.3) that $v_{\rm s} < 175$ km s$^{-1}$. At those relatively low shock velocities, there is no photo-ionized precursor because the speed of the ionization front becomes lower than $v_{\rm s}$; see Equation 4.2 in \cite{dopita96}.} (Equation 3.4 of \citealt{dopita96}): 
\begin{equation}
    n_0/{\mathrm{cm}}^{-3} = \left(10^6 \, {\mathrm{s}}^{3} \, {\mathrm{g}}^{-1}\right) \, L_{\rm{H}\beta}\, A^{-1} \, \left(7.44\, v_2^{2.41}\right)^{-1} 
\end{equation}
where the total area $A$ of the spheroidal bubble can be estimated as $A \approx 2.0 \times 10^{42}$ cm$^2$, and we have already inferred an intrinsic H$\beta$ luminosity $L_{\rm{H}\beta} = (3.2 \pm 0.2) \times 10^{37}$ erg s$^{-1}$ (Section 3.3.5).  This gives $n_0 = (1.3 \pm 0.3)$ cm$^{-3}$, 
where the error includes the uncertainty in the observed flux (estimated as $\delta f_{\rm{H}\beta}/f_{\rm{H}\beta} \lesssim 0.1$) and in the shock velocity ($\delta v_{\rm s}/v_{\rm s} \lesssim 0.2$). So far, we have not discussed the error on the distance $d$ to NGC\,5585. From the standard deviation of redshift-independent distance measurements listed in the NASA/IPAC Extragalactic Database, we infer $\delta d/d \approx 1.5/8$. However, the density scales as $n_0 \propto f_{\rm{H}\beta} \, d^2 \, \Omega^{-1} \, d^{-2} \, v_2^{-2.41} \propto f_{\rm{H}\beta} \, \Omega^{-1} \,  v_2^{-2.41}$, where $\Omega$ is the solid angle of the bubble (an observable quantity); that is, $n_0$ is not directly a function of $d$. For plausible errors $\delta \Omega/\Omega \sim 0.1$, Gaussian propagation shows that the relative error $\delta n_0/n_0$ is dominated by the error on $v_{\rm s}$. We estimate $n_0 = (1.3 \pm 0.6)$ cm$^{-3}$.

The mechanical power is then $P_{\rm w} \approx 1.6 \times 10^{40}$ erg s$^{-1}$, from Equation (4). 
If we adopt instead an intrinsic H$\beta$ luminosity $L_{\rm{H}\beta} \approx 3.4 \times 10^{37}$ erg s$^{-1}$ (corresponding to an intrinsic reddening $E(B-V) \approx 0.02$ mag), the inferred mechanical power increases by the same factor, that is $P_{\rm w} \approx 1.7 \times 10^{40}$ erg s$^{-1}$.
We can estimate the error on $P_{\rm w}$ using similar arguments to those used for the error on $n_0$. In this case, $P_{\rm w} \propto n_0 \, R^5 \, t^{-3} 
\propto f_{\rm{H}\beta} \, d^2 \, v_{\rm s}^{-2.41} \, R^3 \, 
\left(R^{-3} v_{\rm s}^{3}\right) 
\propto f_{\rm{H}\beta} \, d^2 \, v_{\rm s}^{0.59}$, and we obtain $\delta P_{\rm w}/P_{\rm w} \sim 0.4$ within the framework of the assumed model. We conservatively say that our model estimate of $P_{\rm w}$ is valid within a factor of 2.



The second method we used to estimate the jet power is based on the luminosity of suitable diagnostic lines, which carry a known fraction of the total kinetic power \citep{pakull10}. In particular, the fractional power radiated via H$\beta$ and Fe {\footnotesize{II}}$\lambda 1.644\mu$m depends only weakly on the density and shock velocity across our plausible range of parameters. From \cite{dopita96} and the {\sc mappings iii} tables of \cite{allen08}, for Large Magellanic Cloud metallicity (Section 3.3.7) and equipartition magnetic field, we infer that $P_{\rm w} \approx 1/(1.4 \times 10^{-3}) \times L_{\rm{H}\beta} \approx 2.3 \times 10^{40}$ erg s$^{-1}$ (independent of ambient density), in agreement with the previous estimate of the mechanical power. Varying the shock velocity between 100--150 km s$^{-1}$ leads to an error range of $\approx$20\% around the central estimate; choosing a {\sc mappings iii} model without magnetic field leads to an increase in the predicted flux by $\approx$12\%. Correcting the H$\beta$ luminosity for an intrinsic reddening $E(B-V) \approx 0.02$ mag gives $P_{\rm w} \approx 2.4 \times 10^{40}$ erg s$^{-1}$.
If we use the measured Fe {\footnotesize{II}} luminosity, for the same choice of {\sc mappings iii} parameters, we obtain instead $P_{\rm w} \sim 1/(3.4 \times 10^{-4}) \times L_{\rm{[Fe II]}} \approx 3.2 \times 10^{40}$ erg s$^{-1}$, with an uncertainty of $\approx$40\% for the plausible range of shock velocity, and a factor of 2 higher for null magnetic field.



\subsection{Radio luminosity of ULX bubbles}
The integrated radio luminosity is not a particularly reliable proxy for the kinetic power, neither in ULX bubbles nor in radio galaxies, even when the age of the bubble is known. The same kinetic power can produce very different radio luminosities, depending on unobserved quantities such as the filling factor of the magnetic field in the bubble, the minimum and maximum energy of the {\bf {relativistic}} electrons, the relative fraction of power carried by protons. Empirical relations between the optically thin synchrotron luminosity of radio lobes and the jet power were proposed for example by \cite{willott99}, \cite{cavagnolo10}, and \cite{godfrey13}, based on samples of radio galaxies. It would be tempting to extrapolate such relations down to the microquasar regime. However, \cite{godfrey16} argued that such scalings are spurious relations caused by the common dependence on distance of both axes, when the calibration sample is observed over a small range of apparent luminosities but a much larger range of cosmic distances. This introduces a form of Malmquist bias in favour of the most luminous and powerful sources at each distance. See also \cite{feigelson83} and \cite{elvis78} for a discussion of the pitfalls of luminosity-luminosity correlations.

Instead of comparing the size and radio luminosity of this ULX bubble with those of lobes and cavities in radio galaxies, it is perhaps more interesting to compare them with the corresponding quantities in SNRs. An analytical model of the evolution of radio SNRs \citep{sarbadhicary17} predicts 1.4-GHz luminosity density 
\begin{equation}
    L_{1.4}  \approx  6.7 \times 10^{24} \,
    \left(\frac{\epsilon_e}{0.01}\right)\, 
    \left(\frac{\epsilon^u_b}{0.01}\right)^{0.8}\,R_2^3 \,v_2^{3.6}\ 
    {\mathrm{erg~s}}^{-1}\ {\mathrm{Hz}}^{-1} 
\end{equation}
where $R_2$ and $v_2$ were defined in Section 4.1, $\epsilon_e$ is the fraction of kinetic power transferred to relativistic electrons after the shock, and $\epsilon^u_b$ is the fraction of energy in the amplified upstream magnetic field. To a first approximation, $\epsilon_e$ and $\epsilon^u_b$ can be treated as constant \citep{white19}, so that the evolution of the radio luminosity density $L_{1.4}$ depends only on $R(t)$ and $v_{\rm s}(t)$. Let us assume that Eq.(6) holds both for shocked bubbles created by one initial injection of energy $E$ (SNR case) and for those gradually inflated by accretion-powered outflows with constant kinetic power $P_{\rm w}$ (ULX bubbles). Then, the main difference between the two cases is that for an SNR, $R \propto (E/n_0)^{1/5} t^{2/5}$ while for a ULX bubble (as already discussed in Section 4.1) $R \propto (P_{\rm w}/n_0)^{1/5} t^{3/5}$
\citep{pakull06}. 

Inserting the scalings of $R(t)$ and $v_{\rm s}(t)$ for ULX bubbles (Section 4.1), we can recast Equation (6) in these alternative forms, highlighting their dependence of the mechanical power:
\begin{equation}
    L_{1.4}  \approx  6.2 \times 10^{23} \,
    \left(\frac{\epsilon_e}{0.01}\right)\, 
    \left(\frac{\epsilon^u_b}{0.01}\right)^{0.8}\,
    \left(\frac{P_{39}}{n_0}\right)^{1.32}\, t_5^{0.36}\ 
    {\mathrm{erg~s}}^{-1}\ {\mathrm{Hz}}^{-1} 
\end{equation}
\begin{equation}
    L_{1.4}  \approx  1.4 \times 10^{24} \,
    \left(\frac{\epsilon_e}{0.01}\right)\, 
    \left(\frac{\epsilon^u_b}{0.01}\right)^{0.8}\,
    \left(\frac{P_{39}}{n_0}\right)^{1.2}\, R_2^{0.6}\ 
    {\mathrm{erg~s}}^{-1}\ {\mathrm{Hz}}^{-1} 
\end{equation}

For the observed properties of the ULX bubble in NGC\,5585, the predicted radio luminosity density is $L_{1.4} \sim 4 \times 10^{25}$ erg s$^{-1}$ Hz$^{-1}$, within a factor of 3 of the observed luminosity density (Section 3.4). This is a reasonably good agreement, considering the uncertainty on the $\epsilon_e$ and $\epsilon^u_b$ coefficients and the large scatter in the observed radio luminosity of the SNR population. More generally, the synchrotron model of \cite{sarbadhicary17} suggests that a radio luminosity $\sim$10$^{35}$ erg s$^{-1}$ is a characteristic order-of-magnitude value for ULXs with kinetic power of $\sim$10$^{40}$ erg s$^{-1}$ and activity age $\sim$ a few 10$^5$ yr.

It is useful to compare Equations (7,8) with the corresponding expressions for an SNR bubble, derived and discussed in \cite{white19}. The radio luminosity of an SNR bubble decreases with time, during the Sedov phase, as $L_{1.4} \propto t^{-0.96}$ or $L_{1.4} \propto R^{-2.4}$. Conversely, an active ULX bubble increases its radio luminosity as it expands, scaling as $L_{1.4} \propto t^{0.36}$ or $L_{1.4} \propto R^{0.6}$. Thus, a radio SNR with a standard energy $E = 10^{51}$ erg can also reach luminosities $\sim$10$^{35}$ erg s$^{-1}$, but only at times $t \lesssim 10^3$ yrs and sizes $\lesssim$10 pc; radio bubbles with sizes $\sim$100 pc and ages of a few 10$^5$ yr can only be powered by super-critical accretion.

\section{Conclusions}
We showed that the large ionized nebula at the outskirts of NGC\,5585 is one of the cleanest examples of shock-ionized ULX bubbles. Its size (350 pc $\times$ 220 pc, from {\it HST} narrow-band images) puts it in the same class as the bubbles around Holmberg IX X-1 \citep{pakull06,moon11}, NGC\,7793-S26 \citep{pakull10,soria10}, and NGC\,1313 X-2 \citep{pakull02,weng14}. There is no direct evidence of hot spots or of a jet (unlike in NGC\,7793-S26), but we discovered a very strong radio emission, spatially coincident with the optical bubble, with resolved internal structure. In fact, its radio luminosity is slightly higher than in NGC\,7793-S26 \citep{soria10}; it is ten times more radio luminous than Holmberg IX X-1 \citep{berghea20} and two orders of magnitude more luminous than NGC\,1313 X-2 (undetected in the radio).

Using {\it HST} images and LBT spectra, we resolved the ionized nebula into a main ULX bubble and a smaller H {\footnotesize{II}} region apparently located at its northern end. 
We cannot determine whether the ULX bubble and the H {\footnotesize{II}} region are physically connected or their proximity is due to a projection effect. If physically connected, the other intriguing but for the time being unanswerable question is whether the formation of a cluster of young stars at the northern tip of the ULX bubble may have been triggered by the shock wave itself (perhaps even by a jet), or is just a coincidence. 

The ULX bubble is completely dominated by collisional ionization, with no evidence of X-ray photo-ionization; this is inferred in particular from the low [O {\footnotesize{III}}]$\lambda 5007$ and He {\footnotesize{II}}$\lambda 4686$ emission compared with other bubbles, and relatively high 
[Fe {\footnotesize{II}}]$\lambda 1.64 \mu$m, 
[S {\footnotesize{II}}]$\lambda6716,6721$, 
[O {\footnotesize{I}}]$\lambda6300$.
The average shock velocity (inferred from the FWHM of the optical emission lines), assumed to be essentially identical to the bulk expansion velocity of the bubble, is $\approx$100--125 km s$^{-1}$ (with likely higher velocities up to $\approx$150 km s$^{-1}$ along the north-south direction). This is intermediate between the slower shocks seen in NGC\,1313 X-2 ($\approx$80 km s$^{-1}$: \citealt{pakull02,weng14}) and the faster shocks of NGC\,7793-S26 ($\approx$250 km s$^{-1}$: \citealt{pakull10}). Consequently, the estimated dynamical age ($\approx$6 $\times 10^5$ yr) is intermediate between the younger NGC\,7793-S26 ($\approx$2 $\times 10^5$ yr) and the older NGC\,1313 X-2 ($\approx$8 $\times 10^5$ yr).

We used three alternative methods (standard bubble theory, H$\beta$ luminosity and Fe {\footnotesize{II}} luminosity) to estimate the long-term-average mechanical power that is inflating the bubble. We found a consistent result of $\sim$2 $\times 10^{40}$ erg s$^{-1}$. This is higher than the current X-ray luminosity of the central source NGC\,5585 X-1, as was already noted in other ULX bubbles such as NGC\,7793-S26.

The X-1 source itself is by no means an extreme or peculiar example of ULXs. From two {\it XMM-Newton} and two {\it Chandra} observations, we determined a mildly variable luminosity of $\approx$2--4 $\times 10^{39}$ erg s$^{-1}$ (assuming isotropic emission), and a spectrum reasonably well fitted with a disk model with peak temperature $kT_{\rm in} \sim$1.3--1.5 keV, inner radius $\sim$100 km, and low intrinsic absorption ($\lesssim 10^{21}$ cm$^{-2}$). A standard disk-blackbody fit is slightly improved (just at the 90\% confidence level) when we add a (weak) Comptonized power-law tail and/or when we allow the radial temperature index $p$ to be $<$0.75 (slim disk scenario). Both corrections have the same effect of making the X-ray spectrum slightly less curved than a standard disk. This type of broadened disk spectrum is often seen in ULXs that are within a factor of 3 above their Eddington luminosity.

We showed that there is a blue, point-like optical counterpart at the X-ray position of X-1, standing out from the other stellar objects in the surrounding few arcsec. The absolute magnitude of this source ($M_V \approx -4.3$ mag) is also very typical of ULXs: it is consistent either with a young, massive donor star ($M_{\ast}/M_{\odot} \approx 20 \pm 5$ if it is a main sequence or subgiant star) or with optical re-emission from the outer rings of an X-ray irradiated disk.

In conclusion, with this study, we have added another interesting specimen to the class of shock-ionized nebulae around super-critical accreting X-ray binaries. It is now almost two decades since their nature and importance as a diagnostic of mechanical output power were first recognized \citep{pakull02}. Over this time, we have discovered several new cases, and we have made quantitative progress in their interpretation, refining the derivation of their age and kinetic power from their optical/infrared images and spectra. In forthcoming studies, we will focus on the search for shock-ionized bubbles without an associated central X-ray source: this will constrain the beaming factor and the duty cycle of super-critical accreting sources.


\section*{Acknowledgements}
RS acknowledges support and hospitality from the Curtin Institute of Radio Astronomy (Perth, Australia) and from the Observatoire de Strasbourg during part of this work. We thank the anonymous referee for their careful reading of the first version of this paper, and their insightful comments and suggestions. We also thank William Blair, Jifeng Liu, Knox Long, Thomas Russell, Paul Sell, Richard White, Frank Winkler for discussions about SNRs and ionized bubbles, and Robert Fesen for the spectral data files from the 1994 Hiltner telescope observations. This paper uses data taken with the MODS spectrographs built with funding from NSF grant AST-9987045 and the NSF Telescope System Instrumentation Program (TSIP), with additional funds from the Ohio Board of Regents and the Ohio State University Office of Research. We used {\sc iraf} software for optical analysis: {\sc iraf} is distributed by the National Optical Astronomy Observatory, which is operated by the Association of Universities for Research in Astronomy (AURA) under a cooperative agreement with the National Science Foundation. The National Radio Astronomy Observatory is a facility of the National Science Foundation operated under cooperative agreement by Associated Universities, Inc. This research has made use of data obtained from the 3XMM {\it XMM-Newton} serendipitous source catalogue compiled by the 10 institutes of the {\it XMM-Newton} Survey Science Centre selected by ESA. 

\section*{Data Availability}
The {\it Chandra}, {\it XMM-Newton}, {\it HST} and {\it VLA} datasets used for this work are all available for download from their respective public archives. The {\it LBT} data can be provided upon request.

\appendix

\section{Line width and shock velocity}

Inferring a shock velocity from the width of the collisionally ionized optical emission lines is not straightforward, but we can make the following quantitative argument (based on \citealt{dewey10}) with some simplifying assumptions. First, we must distinguish between adiabatic shocks and fully radiative shocks. In the former case, the gas behind the shock picks up a bulk velocity $v_{\rm bulk} = (3/4)\,v_{\rm s}$ ({\it i.e.}, it lags the shock); in the latter case, the shocked gas moves with the shock, $v_{\rm bulk} = v_{\rm s}$, and gets swept up in a thin shell. The analysis of \cite{dewey10} is mostly focused on adiabatic shocks, in which the post-shock equilibrium temperature is $T \approx (3/16)\, (1/k)\, \mu m_{\rm p} v^2_{\rm s} \approx 3 \times 10^5 v^2_2$ K. Instead, for ULX bubbles, we assume here that it is more appropriate to use the fully radiative shock approximation. Our assumption is justified by the fact that the shocks are not very young, and most of the shocked gas has already recombined and cooled to a range of temperatures ($T \sim 10,000$--$50,000$ K) where it emits low ionisation forbidden lines, such as those we observe from [O {\footnotesize {I}}], [S {\footnotesize {II}}], [N {\footnotesize {II}}] and [Fe {\footnotesize {II}}]. In practice, though, we do expect some radiating material also with velocities $(3/4)\,v_{\rm s} \lesssim v_{\rm bulk} \lesssim v_{\rm s}$.

Next, we need to estimate the expansion velocity of the bubble along our line of sight. If we observe the projected central region of a spherical bubble, we see only shocked material coming towards us at $v \approx v_{\rm bulk}$ (front side of the thin bubble) and material receding at $v \approx - v_{\rm bulk}$ (back side of the bubble). This corresponds to a velocity distribution function in the central region 
$f_{\rm c}(v) = (1/2)\, [\delta(v - v_{\rm {bulk}}) + \delta(v + v_{\rm {bulk}})]$, where we have used two delta functions for the two projected velocity components, and a mean velocity $\overline{v} = 0$. From the velocity distribution, we can then calculate the variance $\sigma^2_{\rm c} \equiv \int v^2 \, f_{\rm c}(v) \, dv - (\overline{v})^2 = v_{\rm {bulk}}^2$. Hence, the root-mean-square $\sigma_{\rm c} = v_{\rm {bulk}} = v_{\rm {s}}$, and the Gaussian FWHM observed from the central region is FWHM$_{\rm c} \equiv 2\sqrt{\ln 4} \, \sigma_{\rm c} \approx 2.355 \, v_{\rm {s}}$. Finally, we obtain 
\begin{equation}
v_{\rm {s}} \approx 0.425 \, {\rm {FWHM}}_{\rm c}.
\end{equation}
In practice, the measured value of FWHM$_{\rm c}$ will be a little higher than predicted by the $\delta$-function approximation, both because we are observing a finite (not point-like) region across the centre of the bubble, and because of the likely presence of some cooling gas with velocities $(3/4)\,v_{\rm s} \lesssim v_{\rm bulk} \lesssim v_{\rm s}$. From our analysis of long slit spectra for another shock-ionized bubble, S26 in NGC\,7793 (M.W. Pakull et al., in prep.; see also \citealt{pakull10}), we estimate $v_{\rm {s}} \approx 0.47 \, {\rm {FWHM}}_{\rm c}$. In the adiabatic shock approximation (all the gas at $v_{\rm bulk} = (3/4)\,v_{\rm s}$), the relation between the fitted FWHM$_{\rm c}$ and the shock velocity is $v_{\rm {s}} \approx 0.57 \, {\rm {FWHM}}_{\rm c}$ \citep{dewey10}, which we can take as an upper limit for our estimate of $v_{\rm {s}}$.

Instead of measuring only the lines from central region, we can take spectra for the total emission of the bubble, approximated again as a thin spherical shell with uniform expansion speed of $v_{\rm {bulk}}$. In this case, as argued by \cite{dewey10}, we would observe a (total) uniform distribution function $f_{\rm T}(v) = 1/(2v_{\rm {bulk}}) = {\mathrm{const}}$ in radial velocity space, {\it i.e.}, constant from $v = -v_{\rm {bulk}}$ to $v = +v_{\rm {bulk}}$. Such a distribution has a mean velocity $\overline{v} = 0$. The variance is $\sigma^2_{\rm T} \equiv \int v^2 \, f_{\rm T}(v) \, dv - (\overline{v})^2 = (1/3)\, v_{\rm {bulk}}^2 = (1/3)\, v_{\rm {s}}^2$. Hence, the root-mean-square $\sigma_{\rm T} = (1/\sqrt{3})\,v_{\rm {s}}$, and the Gaussian width is FWHM$_{\rm T} \equiv 2\sqrt{\ln 4} \, \sigma_{\rm T} \approx 1.36 \, v_{\rm {s}}$. Finally, we obtain for the full bubble observation
\begin{equation}
v_{\rm {s}} \approx 0.735 \, {\rm {FWHM}}_{\rm T}.
\end{equation}
The equivalent expression for an adiabatic shock is $v_{\rm {s}} \approx 0.98 \, {\rm {FWHM}}_{\rm T}$ \citep{dewey10,heng10}.

In our specific case, we do not have an integral field spectroscopic measurement of the emission from the whole bubble. However, we do have long-slit spectra approximately oriented along the major and minor axes. We can compute an average of the observed Gaussian widths (corrected for instrumental broadening) for a sample of strong lines extracted from the central section of the two slits, as a proxy for FWHM$_{\rm c}$. Moreover, we can compute an average intrinsic width of the strongest lines over the whole length of the two slits (obviously excluding the H {\footnotesize{II}} region in the Obs1 slit), as a proxy for FWHM$_{\rm T}$.


\bsp	
\label{lastpage}

\end{document}